%% file: paper.tex
\documentclass[12pt]{article}

\include{_preamble}

\title{\titlepaper}

\begin{document}
\maketitle

\begin{abstract}
The presence of intermediate confounders, also called recanting witnesses, is a fundamental challenge to the investigation of causal mechanisms in mediation analysis, preventing the identification of natural path-specific effects. Proposed alternative parameters (such as  randomizational interventional effects) are problematic because they can be non-null even when there is no mediation for any individual in the population; i.e., they are not an average of underlying individual-level mechanisms. In this paper we develop a novel method for mediation analysis in settings with intermediate confounding, with guarantees that the causal parameters are summaries of the individual-level mechanisms of interest. The method is based on recently proposed ideas that view causality as the transfer of information, and thus replace recanting witnesses by draws from their conditional distribution, what we call ``recanting twins". We show that, in the absence of intermediate confounding, recanting twin effects recover natural path-specific effects. We present the assumptions required for identification of recanting twins effects under a standard structural causal model, as well as the assumptions under which the recanting twin identification formulas can be interpreted in the context of the recently proposed \textit{separable effects} models. To estimate recanting-twin effects, we develop efficient semi-parametric estimators that allow the use of data driven methods in the estimation of the nuisance parameters. We present numerical studies of the methods using synthetic data, as well as an application to evaluate the role of new-onset anxiety and depressive disorder in explaining the relationship between gabapentin/pregabalin prescription and incident opioid use disorder among Medicaid beneficiaries with chronic pain.
\end{abstract}

\section{Introduction}
Causal mediation analyses are a set of statistical techniques that can be used to study the mechanisms via which exposures exert causal effects on outcomes. While approaches to mediation analysis vary in how they achieve their goal, a common approach is to start out with the total effect of an exposure on an outcome of interest, and aim to tease out the effect operating through a candidate mediator and/or an effect operating independently of it. Seminal work on mediation analysis includes the early work of \cite{wright1934method} on path analysis and the structural equation methods of \cite{baron1986moderator}. While these methods spurred much progress in applications, they were limited to linear regression models and were not defined within a rigorous causal inference framework that would unveil the assumptions required for their correctness. As a solution to these limitations, \cite{robins1992, Pearl01} generalized the framework to non-parametric models using causal diagrams and counterfactual logic, resulting in the definition of the familiar natural direct and indirect effects (henceforth NDE and NIE). A key realization is that, in a non-parametric structural equation causal model (NPSEM), the NDE and NIE estimands are identifiable from the observed data when (i) all confounders of the exposure-outcome, exposure-mediator and mediator-outcome relationships are measured, and (ii) none of the mediator-outcome confounders are themselves affected by the exposure, i.e., there are no intermediate confounders of the mediator-outcome relation \citep{avin2005identifiability}. Unfortunately, intermediate confounders are pervasive in applied research, which poses an important limitation to wide applicability of natural direct and indirect effects.

Multiple methods to address intermediate confounding  have been proposed in the literature. Many of these
methods focus on estimating path-specific effects that are defined by nested counterfactuals \citep{vanderweele2014effect, steen2017medflex, mittinty2020longitudinal, vo2022longitudinal}. For instance, one can assess the indirect effect mediated via the path that starts from the exposure and goes directly to the mediator without involving the intermediate confounder.  Unfortunately, not all path-specific effects can be identified from the observed data, due to the intermediate confounder being a so-called \textit{recanting witness} \citep{avin2005identifiability}, i.e., a counterfactual variable that tells one story for purposes of one path but tells a different story for purposes of other paths.  As an alternative, one may focus on randomized interventional effects, which consider setting the mediator at a value randomly drawn from the mediator's counterfactual distribution under a particular exposure level \citep{vanderweele2014effect, vansteelandt2017interventional, zheng2017longitudinal}. An interventional indirect effect via a mediator hence captures the combined effect along all underlying causal pathways leading from exposure (possibly via the intermediate confounders and other mediators) to the mediator of interest, then from the mediator directly to the outcome.  Interventional effects, however, fail to provide a decomposition of the average exposure effect. Besides, recent developments show that in certain settings, interventional effects do not satisfy the sharp null criterion for indirect effect measures \citep{miles2022causal}. As a consequence, an interventional indirect effect can be non-null even when there is no individual-level indirect effect.

An often cited related challenge is that defining the NDE and NIE requires consideration of counterfactual variables defined as the hypothetical outcome in a world where an individual is given the exposure of interest, but the mediator is assigned as the value it would have taken under no exposure, so called cross-world counterfactuals. Proponents of an agential view of causality, whereby causal effects must be defined with respect to a feasible action that would elicit them, take issue with this definition as it is impossible to conceptualize a feasible intervention that would yield cross-world counterfactual outcomes (see \cite{diaz2022causal} for further discussion on this point). Furthermore, identification of the NDE and NIE require certain independencies between cross-world counterfactuals, which cannot hold in the presence of intermediate confounders \citep{andrews2020insights, diaz2020causal}.  In order to avoid working with cross-world counterfactual outcomes, a recent strand of the literature on mediation \citep{robins2010alternative, robins2022interventionist, stensrud2021generalized, stensrud2022conditional} considers so called \textit{separable effect} models, which assume that the effect of the exposure on downstream variables works through independent mechanisms. Because these independent mechanisms can be potentially intervened upon, natural direct and indirect effects can be defined in terms of feasible interventions on these mechanisms, at least in principle. 

As a solution to the above challenges, \cite{diaz2022causal} outlines a proposal similar to randomized interventional effects where, instead of randomizing the mediator of interest, the path-specific effects are defined in terms of random draws from the distribution of the recanting witness. In this paper, we use the term \textit{recanting twin}\footnote{We would like to thank Eric Tchetgen Tchetgen for  suggesting this name in an informal conversation.} to refer to these random draws. The resulting causal effects address some of the above issues and have the following properties: (a) allow one to decompose the total average causal effect into path-specific effects, (b) the
path-specific effects satisfy appropriately defined sharp null
criteria, and provide an additional measure of the strength of the intermediate confounding effect. However, the effects proposed in \cite{diaz2022causal} are defined with respect to cross-world counterfactuals, and therefore do not conform to an agential view of causality. In addition, the relationship between path-specific effects defined by recanting twin and their (non-identifiable) natural analogues, as well as the estimation of recanting twin effects have not been formally elaborated.

In this paper, we aim to further extend the concept of recanting twins in multiple directions. First, we investigate recanting twins under models where (a) the effects of the exposure are separable and (b) the effects of both the exposure
and the intermediate mediator are separable, and show that while
identification in model (a) still requires some cross-world independence assumptions, identification under model (b) is achieved
without cross-world assumptions. This results in path-specific effects
that provide falsifiable evidence in the sense that a randomized study can be developed, at least in principle, where the effects are identified under no assumptions. Second, we show that the path-specific effects defined by recanting twins and their natural analogues are equivalent in the absence of recanting witness. We propose a test of the null hypothesis of no intermediate confounding effect by post-baseline variables preceding the mediator of interest. When we fail to reject the null hypothesis, an interpretation of recanting-twin effects as natural effects may be appropriate. In contrast, when the test is rejected, the
two types of effects are not, but a fine-grained decomposition of the total causal effect into different recanting-twin path-specific components
is still attainable from the observed data. This is a major advantage of recanting twin effects compared to (non-identifiable) natural effects. Finally, we propose an efficient semi-parametric estimator of recanting-twin effects, which achieves $\sqrt{n}$-rate of convergence to the parameters of interest, even when the nuisance functions are estimated at slower rates, e.g. by using flexible, data-adaptive or machine learning methods.  

The manuscript is organized as follows. In section 2, we introduce three structural causal models under which the problem of intermediate confounding will be discussed. We then formalize in section 3 the notion of recanting twin and explain how it can be used to address intermediate confounding. In section 4 and 5, we discuss the identification and estimation of path-specific effects defined by recanting twin under different structural causal models. We evaluate the proposed approach by simulated data in section 6, then apply to Medicaid claims data to investigate the role of new-onset anxiety and depressive disorder in explaining the relationship between gabapentin/pregabalin prescription and incident opioid use disorder among patients with chronic pain conditions who have been prescribed opioids, taking into account intermediate confounding by benzodiazepine co-prescriptions and the number of opioid providers (section 7). Section 8 concludes. 

\section{Preliminaries}
Our aim is to investigate the role of a mediator $M$ in explaining the causal relationship between a binary exposure $A$ and an outcome $Y$, in the presence of some common causes $Z$ of $M$ and $Y$ that are affected by $A$. We let $W$ denote a set of baseline covariates, and assume access to observations $O_1,\ldots, O_n$ which are $n$ independent and identically distributed copies of $O=(W,A,Z,M,Y)$. Without loss of generality, we further assume that all densities of $O$ are positive to avoid positivity violations \citep{petersen2012diagnosing}. 

  \label{fig:dag}  
  We formalize our definitions of causality in a structural causal
  model (SCM), also known as nonparametric structural equation
  model \citep{Pearl01, bongers2021foundations}. An SCM is a generative
  model that assumes the existence of unknown but deterministic
  functions that are used to assign the value of each random variable
  as a function of the past variables, where stochasticity is modelled
  through the probability distribution of exogenous random
  variables. The SCM corresponding to our application is given in
  Definition~\ref{def:scm1}, and its associated causal directed
  acyclic graph (DAG) is given in Figure~\ref{fig:dag}.

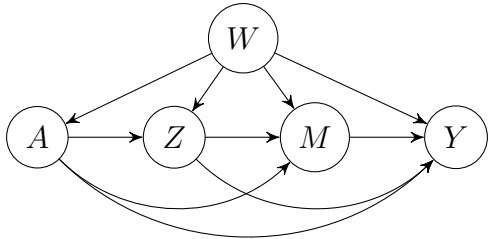
\begin{figure}[!htb]
  \centering
  \begin{tikzpicture}
    \tikzset{line width=1pt, outer sep=0pt,
      ell/.style={draw,fill=white, inner sep=2pt,
        line width=1pt}};
    \node[circle, draw, name=a, ]{$A$};
    \node[circle, draw, name=z, right = 10mm of a]{$Z$};
    \node[circle, draw, name=m, right = 10mm of z]{$M$};
    \node[circle, draw, name=y, right = 10mm of m]{$Y$};
    \node[circle, draw, name=w, above right = 7mm and 3mm of z]{$W$};

    \draw[->](a) to (z);
    \draw[->](w) to (z);
    \draw[->](z) to (m);
    \draw[->](a) to[out=-45,in=225] (y);
    \draw[->](a) to[out=-45,in=225] (m);
    \draw[->](z) to[out=-45,in=225] (y);
    \draw[->](w) to (y);
    \draw[->](m) to (y);
    \draw[->](w) to (a);
    \draw[->](w) to (m);
  \end{tikzpicture}
  \caption{Causal DAG associated to model $\mathcal{M}_1$.}
  \label{fig:dag}  
\end{figure}

\begin{definition} \label{def:scm1} Let
  $\mathcal{M}_1=\langle U, X, f, \mathcal{P}\rangle$ be a SCM, where
  $X=(W,A,Z,M,Y)$ are the endogeneous variables,
  $U = (U_W, U_A, U_Z, U_M, U_Y) \sim \P_U$ are the exogeneous
  variables, $\mathcal{P}$ is the set of allowed distributions $\P_U$,
  and $f = (f_W, f_A, f_Z, f_M, f_Y)$ are deterministic functions such
  that:
    \begin{gather*}
        W= f_W(U_W); \qquad  A = f_A(W, U_A); \qquad Z = f_Z(W, A, U_Z);\\
        M = f_M(W, A, Z, U_M); \qquad  Y = f_Y(W, A, Z, M, U_Y). 
    \end{gather*}
    We call $\mathcal{M}_1$ a standard SCM.
\end{definition}
The proposed functional characterisation in $\mathcal{M}_1$ allows us
to specify how the distribution of the observed data would change in
response to external interventions on certain endogeneous variables
\citep{pearl1995causal}. Here, we use the word \textit{intervention} to refer to a mathematical operation, without regards to whether this mathematical operation corresponds to actions that can be carried out by an agent in the real world \citep[for more discussion on this point see][]{diaz2022causal}. More specifically, denote
$(Z(a), M(a), Y(a))$ the counterfactual values of $(Z,M,Y)$ that could
be observed under an intervention that sets $A$ to $a\in
\{0,1\}$. These counterfactuals are generated by replacing the
structural equation $A = f_A(W, U_A)$ in $\mathcal{M}_1$ by the
degenerate function $A=a$, but keeping the system otherwise unchanged,
that is:
    \begin{gather*}
        W= f_W(U_W); \qquad  A = a; \qquad Z(a) = f_Z(W, a, U_Z);\\
        M(a) = f_M(W, a, Z, U_M); \qquad  Y(a) = f_Y(W, a, Z(a), M(a), U_Y). 
    \end{gather*}
    One can also define other counterfactuals in a similar
    fashion. Specifically, under an intervention that sets a subset of
    endogeneous variables $X_I$ in $X$ to a specific value $x_I$, we
    denote $X_j(x_I)$ the potential value of the non-intervened
    variable $X_j\in X\setminus X_I$.  Restrictions on the
    distribution $\P_U \in \mathcal{P}$ of the exogeneous variables
    $U$ can then be imposed to imply independencies between
    certain counterfactuals. This will in turn allow the
    identification of the causal effects of interest. These assumptions will be discussed in the next
    section.

\subsection{Separable effect models}

In this paper we also develop results for alternative SCMs corresponding to models with separable effects \citep{robins2010alternative,
  robins2022interventionist,stensrud2021generalized,stensrud2022conditional}. These models additionally assume that
exposure $A$ can be separated into three different components $A_Z$,
$A_M$, and $A_Y$, such that the indirect and direct effect of $A$ on
$M$ is solely attributed to $A_Z$ and $A_M$, respectively, and the
direct effect of $A$ on $Y$ (not via $M$ and $Z$) is solely attributed
to $A_Y$. In the current dataset, patients are nonetheless only
exposed to either $A=1$ (equivalent to $A_M=A_Y=A_Z=1$) or $A=0$
(equivalent to $A_M=A_Y=A_Z=0$). The different components of $A$ are
thus not distinguishable from data and must be assumed as part of the
causal model, that is $A \equiv A_M \equiv A_Y \equiv A_Z$.
\begin{definition}\label{def:se}
  Let $\mathcal{M}_2=\langle U, X, f, \mathcal{P}\rangle$ denote the
  SCM associated with endogeneous variables $X=(W,A,Z,M,Y)$,
  exogeneous variables $U = (U_W, U_A, U_Z,$ $U_M, U_Y)$ and
  deterministic functions $f = (f_W, f_A, f_Z, f_M, f_Y)$ such that:
    \begin{gather*}
        W= f_W(U_W); \qquad  
        A= f_A(W, U_A);\qquad
        Z = f_Z(W, A_Z(A), U_Z);\\
        M = f_M(W, A_M(A), Z, U_M); \qquad  Y = f_Y(W, A_Y(A), Z, M, U_Y),
    \end{gather*}
    where $A_Z$, $A_M$, and $A_Y$ are deterministic, known
    transformations or components of the
    exposure. 
    We call $\mathcal{M}_2$ a SCM with separable exposure effects.
\end{definition}
In an abuse of notation, the above definition uses $f$ to denote the
structural equations, but these are not assumed to be the same
functions as in Definition~\ref{def:scm1}. Note that $\mathcal{M}_2$
and $\mathcal{M}_1$ are identical in principle since one can consider,
for example, compositions of $f_Z$ and $A_Z$ to recover
Definition~\ref{def:scm1}. However, $\mathcal{M}_2$ is a subset of
$\mathcal{M}_1$, due to the restriction on the existence of functions $A_Z$, $A_M$, and
$A_Y$. The additional value of $\mathcal{M}_2$ lies in its
capability of isolating the components of $A$ that impact $Z$, $M$,
and $Y$, respectively, and in conceptualizing hypothetical but
feasible interventions on those components that would allow the study of path-specific effects. To illustrate this point, we will consider the following
example, which is an extension of the example given in Section 5.1 of
\cite{robins2010alternative}.
Let $A$ denote smoking status, $M$ denote hypertensive status, $Z$
denote vasopressin, and $Y$ denote whether a patient had a myocardial
infarction. Let $A_Z(A)$ whether nicotine is delivered to the
hypothalamus (where vasopressin is synthesized), let $A_M(A)$ denote
whether nicotine is released to all other organs, and let $A_Y(A)$
denote the presence of all other cigarette toxins. Under these
definitions, the model assumes that the entire effect of nicotine on
myocardial infarction is through its effect on hypertension status,
while the non-nicotine toxins in cigarettes have no effect on
hypertension. Furthermore, it assumes that the entire effect of
nicotine in the hypothalamus on myocardial infarction is through the
release of vasopressin and not through the release of other hormones
or functions of the hypothalamus. Defining path-specific effects using
an extension of the effects of \cite{robins2010alternative} to this
problem will require considering interventions on $A_Z$, $A_M$, and
$A_Y$, rather than interventions on $A$ and $M$.

In the observed data, there is a deterministic relation between $A$
and $A_Z$, $A_M$, and $A_Y$: the smoking status of a patient
determines whether nicotine is released to the hypothalamus and to all
other organs, and whether other cigarettes toxins are present.
Importantly, the assumption of separability can be refuted by
hypothetical future experiments
\citep{robins2022interventionist,stensrud2022separable}. For example,
consider a hypothetical multifactorial trial where patients are
randomized to receive or to not receive each combination of the three
components $A_Z$, $A_M$ and $A_Y$, and let $E_G$ denote expectation
under this
experiment. 
Under separability of exposure effects according to model
$\mathcal{M}_2$, $E_G(Z|A_Z=a_Z, A_M=a_M, A_Y=a_Y)$ would be constant
across different values of $a_M$ and $a_Y$. Likewise,
$E_G(M|A_Z=a_Z, A_M=a_M, A_Y=a_Y)$ would be constant across different
values of $a_Y$.

We note that the definition of separable effects in Definition \ref{def:se} is
different from the original definition given by
\cite{robins2010alternative} in that we use a structural causal model
as the primitive, whereas they use
counterfactuals as the primitive. In a model with no assumptions on
the distribution of the errors $U$, the two definitions are
equivalent.


In this work, we will consider a third model in which the effects of
the intermediate variable $Z$ are also assumed to be
separable. 
\begin{definition}
  Let $\mathcal{M}_3=\langle U, X, f, \mathcal{P}\rangle$ denote the
  SCM associated with endogeneous variables $X=(W,A,Z,M,Y)$,
  exogeneous variables $U = (U_W, U_A, U_Z,$ $U_M, U_Y)$ and
  deterministic functions $f = (f_W, f_A, f_Z, f_M, f_Y)$ such that:
    \begin{gather*}
        W= f_W(U_W); \qquad  
        A = f_A(W, U_A);\qquad
        Z = f_Z(W, A_Z(A), U_Z);\\
        M = f_M(W, A_M(A), Z_M(Z), U_M); \qquad  Y = f_Y(W, A_Y(A), Z_Y(Z), M, U_Y),
    \end{gather*}
    where $A_Z$, $A_M$, and $A_Y$ are deterministic, known
    transformations or components of the exposure, and $Z_M$, $Z_Y$
    are deterministic, known transformations or components of the
    intermediate confounder. We call $\mathcal{M}_3$ a SCM with
    separable exposure effects and separable intermediate confounders
    effects.
\end{definition}

Continuing with our smoking example, $Z_M$ could denote the
vasopressin released in the circulatory system, which increases blood
pressure through vasoconstriction, and $Z_Y$ could denote the
vasopressin released directly into the brain, which has been linked to
social behaviors that can increase stress and affect the likelihood of
a myocardial infarction. As before, $\mathcal M_3$ makes more
assumptions about the data generating mechanism and is therefore a
submodel of $\mathcal M_2$ and $\mathcal M_1$.

To falsify the assumption of separability of the effects of $Z$, we
can conduct a hypothetical trial that randomly assigns each patient to
each combination of the components $Z_M$ and $Z_Y$, and then tests
whether $E_G(M|Z_M=z_M, Z_Y=z_Y)$ is constant across $z_Y$, where
$E_G$ denotes expectation in this hypothetical
experiment. 

In what follows, we will discuss the identification of path-specific
effects in the context of models $\mathcal M_1$, $\mathcal M_2$, and
$\mathcal M_3$. Because these models make increasingly more
assumptions about a-priori knowledge of the data-generating
mechanisms, path-specific effects will naturally be identified under
increasingly weaker
conditions. 

\section{Path-analysis under intermediate confounding}

In this section we introduce the notation $P_j$ to refer to each of
the paths of interest, namely $P_1: A\rightarrow Y$;
$P_2: A \rightarrow Z \rightarrow Y$;
$P_3: A \rightarrow Z \rightarrow M \rightarrow Y$ and
$P_4: A \rightarrow M \rightarrow Y$, where we note that the order we
have chosen is arbitrary.  Analogous to \cite{pearl2001direct},
so-called natural path-specific effects could be defined using the
following nested counterfactuals:
\begin{align}
  Y_{S_0}&=Y(1, Z(1), M(1, Z(1))),\\
  Y_{S_1}&=Y(0, Z(1), M(1, Z(1))),\\
  Y_{S_2}&=Y(0, Z(0), M(1, Z(1))),\label{eq:count} \\
  Y_{S_3}&=Y(0, Z(0), M(1, Z(0))),\\
  Y_{S_4}&=Y(0, Z(0), M(0, Z(0))),
\end{align}
where we simplified notation by using conventions such as
$Y(a,z,m)=f_Y(W, a,z,m,U_Y)$, and where $S_j$ denotes sequential interventions with respect to effects operating through the set of paths $S_j=\{P_1,\ldots,P_j\}$, starting from a reference counterfactual $Y_{S_0}$. The causal effect operating through the path $P_j$ can then be measured as $\E[Y_{S_{j-1}} - Y_{S_j}]$, leading to so-called natural path-specific effects. Note that due to telescoping, the sum of these effects is equal to the average treatment effect
(ATE) $\psi=\E[Y(1) - Y(0)]$. In the SCMs described by $\mathcal M_1$, the distribution of $Y_{S_0}, Y_{S_1} $ and $ Y_{S_3}$ can be nonparametrically identified from the observed data under certain causal assumptions which are satisfied, for example, if all errors $U$ are independent \citep{vanderweele2014effect,  robins2022interventionist}.
Unfortunately, this is not the case for $Y_{S_2}$, due to $Z$ being a \textit{recating witness}. 

To address the non-identifiability of the distribution of $Y_{S_2}$, we propose a novel concept called recanting twin. This concept is formally defined as follows.
\begin{definition}{(Recanting twin)}
  Let $T(a)$ denote a random draw from the distribution of $Z(a)$
  conditional on $W$. For $a\in\{0,1\}$, $Z(a)$ is the recanting witness of $Z(1-a)$, and we say that $T(a)$ is the
  recanting twin of $Z(1-a)$.
\end{definition}
Intuitively, the recanting witness $Z$ prevents the identification of the distribution of $Y_{S_2}$ because $Z(1)$ and $Z(0)$ tell different stories for purposes of the paths $P_2$ and $P_3$. Hence, the identification of $Y_{S_2}$ would require the identification of the joint distribution of $Z(1)$ and
$Z(0)$. To solve this issue, we use the
recanting twin $T(a)$ to ``tell a different story'' instead of $Z(a)$. Specifically, let:
\begin{align*}
  Y_{S_1}' &= Y(0, Z(1), M(1, T(1))),\qquad
  Y_{S_2}'' = Y(0, T(0), M(1, Z(1))),\\
  Y_{S_2}' &= Y(0, Z(0), M(1, T(1))),\qquad
  Y_{S_3}'' = Y(0, T(0), M(1, Z(0))),
\end{align*}
For example, in $Y_{S_2}'$, the effect of the intervention $A=1$ operating through the path $A=1\to Z(1)\to M\to Y$ is emulated by the twin $T(1)$ instead of $Z(1)$, which would have been a recanting witness due to the path $A=0\to Z(0)\to Y$. This allowed \cite{diaz2022causal} to achieve the following identifiable effect decomposition of the average treatment effect $\psi=\E[Y(1) - Y(0)]$ as
\[\psi = \psi_{P_1} + \psi_{P_2} + \psi_{P_3} + \psi_{P_4} +
  \psi_{P_2\vee P_3},\] where
\begin{align*}
  \psi_{P_1} = \E(Y_{S_0} - Y_{S_1}); \quad   \psi_{P_2} = \E(Y_{S_1}' - Y_{S_2}');\quad
  \psi_{P_3} = \E(Y_{S_2}'' - Y_{S_3}'');\quad
  \psi_{P_4} = \E(Y_{S_3}  - Y_{S_4});
  \end{align*}
are the path-specific effects, and:
\[ \psi_{P_2\vee P_3} = E(Y_{S_1} - Y_{S_1}' + Y_{S_2}' - Y_{S_2}'' +
  Y_{S_3}'' - Y_{S_3})\] 
is a parameter measuring the extent of intermediate confounding. Importantly, \cite{diaz2022causal} showed that $\psi_{P_j}$ satisfies path-specific null criteria in the sense that $\psi_{P_j}=0$ whenever there is no individual-level effect through $P_j$. As a result, these effects, if identifiable, can serve as alternative measures for natural path-specific effects when the natural versions are not  identifiable from observed data. 

To shed further insights on the interpretation of $\psi_{P_2\vee P_3}$ and the relation between recanting-twin and natural path-specific effects, we now consider the special setting where $Z$ is not an intermediate counfounder. Such a setting is formalized as follow.
\begin{definition}[No intermediate confounding]\label{def:intconf}
    Let $\mathcal{U}$ denote the range of the random errors $U = (U_W , U_A, U_Z, U_M, U_Y)$ in the SCM. We say that there is no intermediate confounding by $Z$ when $\mathcal{U}$ can be partitioned into subsets $\mathcal{U}_1, \mathcal{U}_2,$ and $\mathcal{U}_3$ such that the following conditions hold almost surely:
\begin{itemize}
    \item $Z(1) = Z(0)$ in $\mathcal{U}_1$, and
    \item sup$_{\overline{z}}\mid M(z_1) - M(z_0) \mid = 0$ in $\mathcal{U}_2$, and
    \item sup$_{\overline{z},m}\mid Y(m,z_1) - Y(m,z_0) \mid = 0$ in $\mathcal{U}_3,$
\end{itemize}
where we let $\overline{z}=(z_1,z_0)$.
\end{definition}
This definition requires that there is no individual in the population for whom the effects $A\rightarrow Z$, $Z\rightarrow M$, and $Z\rightarrow Y$ are all active, to assume away the intermediate confounding effect by $Z$. Remarkably, these effects can be simultaneously non-null on the population level, so the above-mentioned arrows may still co-exist on the causal DAG in Figure~\ref{fig:dag} without violating the condition in Definition \ref{def:intconf}. As a consequence, simply evaluating the absence of each relationship on the population level to judge the absence of intermediate confounding is seemingly too conservative. Besides, it is noteworthy that our definition of no intermediate confounding is refined from what is proposed in \citet{diaz2022causal}, which requires no total effect of $Z$ on $Y$ instead of no direct effect of $Z$ on $Y$ not via $M$ in the subset $\mathcal{U}_3$. The latter requirement is arguably more appropriate, as it can avoid the scenario where the indirect and direct effect (via and not via $M$) of $Z$ on $Y$ add up to zero (in which case $Z$ still acts as an intermediate confounder despite having no total effect on $Y$, given that the arrow $A \rightarrow Z$ is active).

Lemma \ref{lemma:ic} provides  interpretation of $\psi_{P_2\vee P_3}$ and of the recanting-twin effects under no intermediate confounding. A formal proof of this is provided in the Online Supplementary Materials. 
\begin{lemma}\label{lemma:ic}
    When there is no intermediate confounding, $\psi_{P_2 \vee P_3}=\psi_{P_2}\psi_{P_3}=0$ and the path-specific effects defined by recanting twins are equal to their natural analogues, even on an individual level.
\end{lemma}
A consequence of Lemma \ref{lemma:ic} is that the null hypothesis of no intermediate confounding by $Z$ can be evaluated via testing whether $\psi_{P_2 \vee P_3}=0$ or whether $\psi_{P_2}\psi_{P_3}=0$. When these tests are not rejected, an interpretation of recanting-twin effects as natural effects cannot be rejected, due to Lemma \ref{lemma:ic}. When the tests are rejected, the two types of effect are no longer equivalent, and a fine-grained decomposition by natural effects is not possible. In contrast, as is shown in the following sections, recanting twin effects are generally identifiable under different SCMs, assuming different causal assumptions. This then allows one to measure the strength of different causal pathways, and also to measure the strength of the intermediate confounding effect. These properties are major advantages of recanting twin effects compared to their natural analogues.

\section{Identification under different causal structural models}
\subsection{Identification of path-specific effects under no separability}
Under model $\mathcal{M}_1$, let $Y(a,m,z) = f_Y(W, a, z, m, U_Y)$, $M(a,z) = f_M(W, a, z, U_M)$, and $Z(a)=f_Z(W, a, U_Z)$ denote counterfactual variables. We will assume the distribution of the errors $\P_U$ is such that the following assumptions hold.
\begin{assumption}[Sequential ignorability]\label{ass:si1} For all $(a,m,z)$:
\begin{enumerate}[label=(\roman*)]
    \item $Y(a,m,z)\indep A\mid W$; $Y(a,m,z)\indep Z\mid (A=a, W)$; and $Y(a,m,z)\indep M\mid (A=a, Z=z, W)$ 
    \item $M(a,z)\indep A\mid W$; and $M(a,z)\indep Z\mid (A=a, W)$
    \item $(Z(a),M(a))\indep A\mid W$ 
    \end{enumerate}
\end{assumption}

\begin{assumption}[Cross-world counterfactual independence]\label{ass:cw} For all $a,a', a''=0,1$; $z,z'\in \mathrm{supp}(Z)$ and $m\in \mathrm{supp}(M)$:
  \begin{enumerate}[label=(\roman*)]
  \item $Y(a,m,z) \indep (M(a',z'),Z(a''))\mid W;$\label{ass:cw2}
        \item $M(a,z) \indep Z(a')\mid W$.\label{ass:cw1}
        \end{enumerate}
\end{assumption}

A sufficient condition for \ref{ass:si1} and \ref{ass:cw} to hold is that all common causes of any pair of variables among $(A,M,Z,Y)$ are measured and are given by the variables that precede the earliest variable in the causal ordering of the pair. This is formalized in $\mathcal M_1$ as follows:
\begin{assumption}[No unmeasured confounders]\label{ass:nounc}
Assume: 
\begin{enumerate}[label=(\roman*)]
    \item $U_A\indep (U_Y, U_M, U_Z)\mid W$,
    \item $U_Z\indep (U_Y, U_M)\mid (A,W)$, and
    \item $U_M\indep U_Y\mid (Z,A,W)$.
    \end{enumerate}
\end{assumption}

Assumption \ref{ass:nounc} will be sufficient but not necessary for all subsequent identification results in this manuscript. This means that a researcher could ensure that the cross-world assumption is satisfied by ensuring that she has measured all common causes of the relevant variables, without the need to appeal to the metaphysics involved in reasoning about counterfactual independencies. Assumptions \ref{ass:si1} and \ref{ass:cw} are violated when there exist unmeasured common causes of any pair of variables among $(A,M,Z,Y)$. Additionally, assumption \ref{ass:cw} is violated when
there are measured or unmeasured common causes of any two variables among $(Z,M,Y)$ that are themselves affected by the exposure $A$. However, when these so-called ``secondary'' intermediate confounders are measured, a simple solution is to treat them as part of $Z$ to proceed with identification. This is an acceptable strategy whenever the main objective is to assess the mediating role of $M$, so $Z$ is a ``nuisance'' that can include all intermediate confounders.

Finally, note that the cross-world assumption \ref{ass:cw} requires independencies between counterfactuals that are indexed by interventions on both $A$ and $Z$, as opposed to the cross-world independencies required for identification of the natural direct and indirect effects, which require independence between counterfactuals indexed by interventions only on $A$.

Under these assumptions, for $a'=1$ and $a^\star=0$, we obtain the following identification results:
\begin{align}
    \E(Y_{S_0}\mid W) &=\E(Y\mid a', W)\notag\\
        \E(Y_{S_1}\mid W) &=\sum_{z, m}\E(Y\mid a^\star, z, m, W)\cdot \P(z, m\mid a', W)\notag\\
    \E(Y_{S_1}'\mid W) &= \sum_{z,m} \E(Y\mid a^\star,z,m,W) \cdot \P(z\mid a', W) \cdot \P (m\mid a', W)\notag\\
    \E(Y_{S_2}'\mid W) =  \E(Y_{S_2}''\mid W)  &= \sum_{z,m} \E(Y\mid a^\star,z,m,W) \cdot \P(z\mid a^\star, W) \cdot \P (m\mid a', W)\label{eq:iden}\\
    \E(Y_{S_3}''\mid W)  &= \sum_{z,m,z'} \E(Y\mid a^\star,z,m,W) \cdot \P(z\mid a^\star, W) \cdot \P (m\mid a', z',W) \cdot \P(z'\mid a^\star, W)\notag\\
    \E(Y_{S_3}\mid W)  &= \sum_{z,m} \E(Y\mid a^\star,z,m,W) \cdot \P(z\mid a^\star, W) \cdot \P (m\mid a', z,W)\notag\\
        \E(Y_{S_4}\mid W) &=\E(Y\mid a^\star, W)\notag,
\end{align}
from which we can construct identification results for $\psi_{P_j}$ and for $\psi_{P_2\vee P_3}$.


A major and common criticism of path-specific effects defined in terms of cross-world counterfactuals is that their interpretation requires considering infeasible worlds where the exposure can take two distinct values, and their identification requires cross-world independence assumptions that can never be tested nor be made to hold by design. As a result, conclusions arising from such path-specific effects are not falsifiable and are therefore sometimes considered of reduced scientific value. To address this concern, recent efforts have focused on the definition of parameters under so-called separability of effects, which fundamentally states that the effect of a variable on downstream variables can be separated into components that are independent and that may be in principle amenable to experimentation, such as our models $\mathcal M_2$ and $\mathcal M_3$. In what follows, we describe definitions and identification results for effects under separability of the effects of $A$ (model $\mathcal M_2$) and the effects of $A$ and $Z$, (model $\mathcal M_3$). Crucially, although the definition and identification of the causal parameters changes according to the model, the identifying functionals remain identical and equal to those given in Equation (\ref{eq:iden}). This has important implications for practice, as it means that an analyst that proceeds with estimation of these statistical parameters can rely on at least three alternative causal interpretations under different sets of assumptions, some of which may be more valid or useful depending on the specific application.

\subsection{Definition and identification of path-specific effects under separability of treatment effects}

In this section, we consider effects defined in terms of interventions on the random variables $A_Y$, $A_M$, and $A_Z$. When the separability assumptions of $\mathcal M_2$ are correct, these effects will be equal to the effects defined in the previous section, but may be different otherwise. Our goal is to establish the conditions under which the path-specific effects based on recanting twins can be identifiable when the exposure effect is separable, as in model $\mathcal{M}_2$. 

An appropriate definition of the relevant natural counterfactuals in $\mathcal M_2$ in terms of interventions on $A_Y$, $A_M$, and $A_Z$ is as follows:
\begin{align*}
  Y_{S_0}&=Y(1, Z(1), M(1, Z(1))),\\
  Y_{S_1}&=Y(0, Z(1), M(1, Z(1))),\\
  Y_{S_2}&=Y(0, Z(0), M(1, Z(1))), \\
  Y_{S_3}&=Y(0, Z(0), M(1, Z(0))),\\
  Y_{S_4}&=Y(0, Z(0), M(0, Z(0))),
\end{align*}
where, in contrast to the previous section, here we define $Y(a_Y, z_Y, M(a_M, z_M)) = f_Y(W, a_Y, z_Y,\allowbreak f_M(W, a_M, z_M, U_M), U_Y)$, where $Z(a_Z)=f_Z(W, a_Z, U_Z)$. 
As before, path-specific effects can be obtained by contrasting these counterfactual outcomes. However, as before, the distribution of $Y_{S_2}$ is not identified due to the recanting witnesses $Z(1)$ and $Z(0)$. To address this, we can use recanting twins. Let $T(a_Z)$ denote a random draw from the distribution of $Z(a_Z)$ conditional on $W$. The recanting-twin counterfactuals can then be expressed as:
\begin{align*}
    Y_{S_1}' &= Y(1, Z(1), M(0, T(1))),\qquad Y_{S_2}'' = Y(1, T(0), M(0, Z(1))),\\
    Y_{S_2}' &= Y(1, Z(0), M(0, T(1))),\qquad Y_{S_3}'' = Y(1, T(0), M(0, Z(0))),
\end{align*}
Note that, unlike the counterfactuals in the previous section, some of these counterfactuals are not cross-world. 

Under the following assumptions, the expectation of the relevant counterfactuals is identified as in Equation~(\ref{eq:iden}):
\begin{assumption}[Sequential ignorability]\label{ass:si2} For all $(a_Z, a_M, a_Y,m,z)$:
  \begin{enumerate}[label=(\roman*)]
    \item $Y(a_Y,m,z)\indep A\mid W$; $Y(a_Y,m,z)\indep Z\mid (A=a_Y, W)$; and $Y(a_Y,m,z)\indep M\mid (A=a_Y, Z=z, W)$, 
    \item $M(a_M,z)\indep A\mid W$; and $M(a_M,z)\indep Z\mid (A=a_M, W)$,
    \item $(Z(a_Y),M(a_M))\indep A\mid W$,
\end{enumerate}
\end{assumption}
\begin{assumption}[Single-world counterfactual
  independence]\label{ass:sw1} For all $(a_M, a_Z, z, z', m)$:
  \begin{enumerate}[label=(\roman*)]
  \item $\{M(a_M,z) = m\} \indep \{Z(a_Z)=z'\}\mid W$.
  \end{enumerate}
\end{assumption}
\begin{assumption}[Cross-world counterfactual independence]\label{ass:ci1} For all $(a_Z, a_M, a_Y,m,z,z')$:
  \begin{enumerate}[label=(\roman*)]
        \item $Y(a_Y,m,z) \indep (M(a_M,z'),Z(a_Z))\mid W$.
        \end{enumerate}
      \end{assumption}

  As before, the above assumptions are all satisfied if there is no
  unmeasured confounding between any pair of variables in $(A,Z,M,Y)$,
  in the sense of \ref{ass:nounc}. We note that while assumption
  \ref{ass:sw1} is similar in spirit to \ref{ass:cw}\ref{ass:cw1}, it
  is not a cross-world assumption since it can be verified in the single-world intervention graph (SWIG) corresponding to the DAG depicted in Figure \ref{fig:dag}. When assumption \ref{ass:sw1} does not hold but a relaxed version of it does, i.e., $\{M(a_M,z) = m\} \indep \{Z(a_Z)=z\}\mid W$ for all $(a_Z, a_M, a_Y,m,z)$, then the distributions of $Y_{S_2}''$ and $Y_{S_3}''$ (but not $Y_{S_1}'$ and $Y_{S_2}'$) are still identifiable from the observed data (given that assumptions \ref{ass:si2} and \ref{ass:ci1} are satisfied). Note also that this relaxed assumption can 
  hypothetically be
  refuted in an experiment $G$ that randomly assigns $(A_M, A_Z)$ to
  each of its four values, and then tests whether
  $\P_G(M=m\mid Z=z, A_Z=a_Z,A_M=a_M,W=w)$ depends on $z$, where
  $\P_G$ denotes the probability distribution in experiment $G$.

  The separable exposure in model $\mathcal{M}_2$ allows us to reduce
  the complexity of the cross-world assumption in the sense that it
  does not require assumptions on counterfactuals indexed by distinct
  interventions on $A$. However, the remaining cross-world assumptions
  is still generally untestable, even in principle, as it requires to
  consider counterfactual variables indexed by distinct interventions
  on $Z$. To address this, in the next section we will require model
  $\mathcal M_3$, in which the intermediate confounder effects are
  also separable.



\subsection{Definition and identification of path-specific effects under separability of treatment and intermediate confounder effects}
In model $\mathcal{M}_3$, we additionally assume separable
intermediate confounders. An appropriate definition of the relevant
natural counterfactuals in $\mathcal M_3$ in terms of interventions on
$A_Y$, $A_M$, and $A_Z$ is as follows:
\begin{align*}
  Y_{S_0}&=Y(1, Z_Y(1), M(1, Z_M(1))),\\
  Y_{S_1}&=Y(0, Z_Y(1), M(1, Z_M(1))),\\
  Y_{S_2}&=Y(0, Z_Y(0), M(1, Z_M(1))), \\
  Y_{S_3}&=Y(0, Z_Y(0), M(1, Z_M(0))),\\
  Y_{S_4}&=Y(0, Z_Y(0), M(0, Z_M(0))),
\end{align*}
where, in contrast to previous sections, we define
\[Y(a_Y, z_Y, M(a_M, z_M)) =  f_Y(W, a_Y, z_Y,  f_M(W, a_M, z_M, U_M),
U_Y),\] where $Z(a_Z)=f_Z(W, a_Z, U_Z)$; and $Z_Y(a_Z)$ and $Z_M(a_Z)$
are short for $Z_M(Z(a_Z))$, respectively.  As before, we first
re-express the recanting twin counterfactuals by intervening on
different components of the exposure and of the intermediate
confounders. More precisely, let $T_M(a_Z)$ and $T_Y(a_Z)$ denote
random draws from the distribution of $Z_M(a_Z)$ and $Z_Y(a_Z)$
conditional on $W$, respectively. Then,
\begin{align*}
  Y_{S_1}' &= Y(1, Z_Y(1), M(0, T_M(1))),\qquad Y_{S_2}'' = Y(1, T_Y(0), M(0, Z_M(1))),\\
  Y_{S_2}' &= Y(1, Z_Y(0), M(0, T_M(1))),\qquad Y_{S_3}'' = Y(1, T_Y(0), M(0, Z_M(0))),
\end{align*}

The following single-world assumptions on model $\mathcal{M}_3$ will
be sufficient for the relevant path-specific contrasts to be
identified as in Equation (\ref{eq:iden}). A formal proof of this is
provided in Appendix \ref{app:a1}
\begin{assumption}[Sequential ignorability] For all $(a_Z, a_M,
  a_Y,z_M,z_Y,m)$:
  \begin{enumerate}[label=(\roman*)]
  \item $Y(a_Y,z_Y, m)\indep A\mid W$; $Y(a_Y,z_Y,m)\indep Z\mid (A=a_Y, W)$; and $Y(a_Y,z_Y,m)\indep M\mid (A=a_Y, Z=z_Y, W)$, 
  \item $M(a_M,z_M)\indep A\mid W$; and $M(a_M,z_M)\indep Z\mid (A=a_M, W)$,
  \item $(Z(a_Y),M(a_M))\indep A\mid W$,
  \end{enumerate}
\end{assumption}
\begin{assumption}[Single-world counterfactual
  independencies]\label{ass:sw2} For all $(a_Z,a_M, a_Y, z_M, z_Y, m, y)$:
  \begin{enumerate}[label=(\roman*)]
  \item $\{Y(a_Y,z_Y,m)=y\} \indep \{M(a_M,z_M)=m,Z_M(a_Z)=z_M\}\mid
    W$.
  \item $\{Y(a_Y,z_Y,m)=y\} \indep \{M(a_M,z_M)=m,Z_Y(a_Z)=z_Y\}\mid W$.\label{ass:siind2}
  \item $\{M(a_M,z_M)=m\} \indep \{Z_M(a_Z)=z_M\}\mid W$.
  \item $\{M(a_M,z_M)=m\} \indep \{Z_Y(a_Z)=z_Y\}\mid W$.
  \end{enumerate}
\end{assumption}

As before, the assumption of no unmeasured confounders \ref{ass:nounc}
will be sufficient but not necessary for all previous assumptions to
hold. Separability of the effects of $A$ and $Z$ ensures that the
cross-world counterfactual independencies of Assumption \ref{ass:cw}
can be swapped for Assumption \ref{ass:sw2}, which is refutable in an
experiment and therefore not cross-world. For instance, consider the
independence
$\{Y(a_Y,z_Y,m)=y\} \indep \{M(a_M,z_M)=m,Z_Y(a_Z)=z_Y\}\mid W$, and
an experiment $G$ that randomly assigns $A_Z$, $A_M$, $A_Y$, and $Z_M$
(but not $Z_Y$). For instance, in the smoking example, assume it was
possible to randomly assign patients to delivery of nicotine to the
hypothalamus ($A_Z$), to all other organs ($A_M$), smoking a
nicotine-less cigarette ($A_Y$), and vasopressin release in the
circulatory system ($Z_M$). In this experiment we have
$\P(Y(a_Y,z_Y,m)=y\} \mid M(a_M,z_M)=m,Z_Y(a_Z)=z_Y, W=w) = \P_G(Y=y\mid
M=m, Z_Y=z_Y, Z_M=z_M, A_Y=a_Y, A_M=a_M, A_Z=a_Z, W=w)$. Assumption
\ref{ass:sw2}\ref{ass:siind2} can thus be refuted if this function
varies with $m$ or $z_Y$.

Furthermore, the proposed counterfactuals $Y_{S_1}'$, $Y_{S_2}'$,
$Y_{S_2}''$ and $Y_{S_3}''$ are defined with respect to interventions
on $A_Z$, $A_M$, $A_Y$, $Z_M$, and $Z_Y$ only, and not interventions
on $M$. This suggests that the path-specfic effects defined based on
these counterfactuals can have an agential interpretation in situations where interventions on the putative
mediator $M$ are not feasible in the real world
\citep{robins2022interventionist,diaz2022causal}. This, however, does require that
interventions on the intermediate variables $Z_M$ and $Z_Y$ are feasible.


\section{Estimation of path-specific effects based on recanting twins}\label{sec:estima}
In this section, we will discuss the estimation of path-specific effects defined by using recanting twins, based on the identification results provided in (\ref{eq:iden}). In the simplest setting where the mediator $M$ and the intermediate confounders $Z$ are both categorical, it is relatively easy to obtain estimators of the expectations of the recanting-twin counterfactuals given $W$, by postulating and fitting a parametric regression model for the nuisance parameters $\E(Y\mid A, Z, M, W), \P(Z\mid A, W)$ and $\P(M\mid A,Z,W)$. Plugging in these estimates into (\ref{eq:iden}) allows us to obtain estimators of the target parameters that achieve $\sqrt{n}$-rate convergence, under the condition that the proposed parametric regression models are correctly specified. 
However, this approach can induce bias when the nuisance models are incorrectly specified or cannot be described by a finite-dimension vector of parameters. Furthermore, the distribution of the final estimators are generally unknown when model selection, regularization or non-parametric techniques must be employed to estimate the nuisance parameters. 

In view of the above concerns, it is important to develop insights on the first-order bias that arises when using off-the-shelf flexible, data-adaptive or machine learning methods to estimate the nuisance parameters. Adjusting for this bias will allow us to obtain again $\sqrt{n}$-consistent plug-in estimators for the target parameters, under certain conditions on the (slower) rate of convergence of the nuisance estimators \citep{bickel1993efficient, pfanzagl1982contributions, kennedy2022semiparametric}. To characterize this first-order bias, one needs to characterize the efficient influence function (EIF) of the target parameters based on (\ref{eq:iden}). Derivation of the EIF is based on the existence of a function $\phi$ and a residual term $R$ that satisfy the following expansion:
\begin{align}\label{vonmises}
    \theta(\eta^*) - \theta(\eta) = -\E\big\{ \phi(X;\eta^*)\big\} + R(\eta,\eta^*),
\end{align}
where $\eta$ denotes the true value of the nuisance parameters, $\theta(\eta^*)$ denotes the target parameter (i.e. one of the expectations in (\ref{eq:iden})) when plugging in a fixed value $\eta^*$ of the nuisance parameters. If this expansion exists, it will turn out that $\phi$ is the EIF corresponding to $\theta$, and $R(\eta, \eta_1)$ is a second-order term that can be expressed as:
\[R(\eta,\eta^*) = \E\big[ c(\eta, \eta^*)\cdot \{f(\eta^*) - f(\eta)\}\cdot\{g(\eta^*) - g(\eta)\} \big]\]
for functionals $c,f$ and $g$. 

By using the Delta method for functionals, one can then derive the exact formulas of the EIF for each target parameter provided in (\ref{eq:iden}). These formulas and corresponding second-order terms are provided in the Online Supplementary Materials.

Two remarks are noteworthy here. First, when $\eta^*$ equals a preliminary estimator $\hat\eta$ of the nuisance parameters, the second order term $R(\eta,\hat\eta)$ will generally be negligible under some arguably weak requirement on the rate of convergence of the nuisance estimators. For instance, if $\hat \eta$ converges to $\eta$ in $L_2$ norm at rate $n^{-1/4}$ or faster, then the second-order term $R(\eta,\hat\eta)$ is $o_P(n^{-1/2})$, which is ignorable. Second, the remaining bias due to plugging in the nuisance estimate $\hat\eta$ into the target functional $\theta(\cdot)$ will equal $-\E\big\{ \phi(X;\hat \eta)\big\}$.  Estimating this bias by the sample average of $\phi(X_i;\hat \eta)$, and then adding it back to the plug-in estimator will allow us to obtain an updated estimator for $\theta$ that is asymptotically unbiased. More precisely, this (bias-corrected) plug-in estimation when both $Z$ and $M$ are categorical can be implemented as follow. 
\begin{enumerate}
    \item[1.] Estimate $\E(Y\mid A=a, Z, M, W)$ and $\E(Y\mid A=a, W)$ for $a=0,1$, by using a flexible regression method.
    \item[2.] Estimate the conditional probabilities $\P(Z=z\mid A=a, W)$ and $\P(M=m\mid A=a, Z, W)$ for $a=0,1$, by using a flexible regression method.
    \item [3.] Calculate the plug-in estimate $\hat\theta_0$ for each target parameter $\theta$ in (6). 
    \item [4.] Estimate the EIF for each target parameter in (6). 
    \item [5.] Compute the (bias-corrected) plug-in estimator $\hat\theta_1$ for each target parameter $\theta$ in (6), as $\hat\theta_1 = \hat\theta_0 + \frac{1}{n}\sum_{i=1}^n\phi(X_i;\hat\eta)$.
\end{enumerate}  
Of note, the suggested procedure is also applicable to more general and complex settings where $Z$ and $M$ are of high dimension, or include multiple components of different nature, e.g. some components are continuous and some are categorical. In such general settings, the sums over $(z,m)$ or $(z,m,z')$ in (6) are replaced by integrals over the corresponding variables, and the conditional probabilities $\P(z\mid \cdot) $ and  $\P(m\mid \cdot)$ in (6) are replaced by corresponding conditional densities, i.e., $f_Z(z\mid \cdot) $ and  $f_M(m\mid \cdot)$. These densities can be non-parametrically estimated by many flexible, data-adaptive methods proposed in the literature, e.g., \citet{hjort1996locally, fan2004crossvalidation} and \citet{wang2012density}, among many others. The practical challenge, however, lies in the numerical computation of the integrations arising in $\theta$ and the corresponding EIF when $(Z,M)$ are not discrete variables. For now, we will merely focus on the case of categorical $(Z,M)$ to illustrate our proposal, and will address numerical challenges in more general settings in future works. Our semi-parametric development below, nonetheless, will remain valid in general settings.

\begin{theorem}\label{theo1} (Asymptotic normality and efficiency) Assume:
\begin{itemize}
    \item [(i)] The second-order term $R(\eta, \hat\eta)$ is $o_P(n^{-1/2})$ and 
    \item [(ii)] The functions $\phi(X;\eta)$ and $\phi(X;\hat\eta)$ are in a Donsker class. 
\end{itemize}
In that case,
\[\hat\theta_1 = \theta 
+ \frac{1}{n} \sum_{i=1}^n \phi(X_i; \eta) + o_P(1),\]
due to which $\sqrt{n}(\hat\theta_1 - \theta) \xrightarrow{D} N(0,\zeta^2)$, where $\zeta^2 = V\{ \phi(X; \eta)\}$ is the non parametric efficiency bound.
\end{theorem}
As a reminder, condition (i) for asymptotic normality in Theorem \ref{theo1} is satisfied if $\hat{\eta}$ converges in $L_2(P)$ norm to $\eta$ at $n^{-1/4}$-rate or faster. In contrast, condition (ii) (i.e. Donsker condition) may be avoided by using cross-fitting in the estimation procedure \citep{chernozhukov2018double}. To achieve this, the dataset is randomly partitioned into $Q$ sets of approximately equal size, namely $V_1, \ldots , V_Q$. On each sample $T_q = \{1,\ldots,N\} \setminus V_q$, the data-adaptive algorithm will be trained and then used to produce a prediction $\hat{\eta}_{q,i}$ of $\eta$ for each unit $i$ in the validation set $V_q$. The one-step estimator is finally adapted to cross-fitting by substituting all occurrences of $\hat{\eta}(X_i)$ by $\hat{\eta}_{q,i}(X_i)$ in the estimation procedure provided above. 

As a direct consequence of Theorem \ref{theo1}, the variance of the plug-in estimator $\hat\theta_1$ can be estimated by the sample variance of the efficient influence function, i.e. $\hat\zeta^2 = \hat V\{\phi(X,\hat{\eta})\}$, with $\hat\theta_1$ and the nuisance parameter vector $\eta$ estimated as in the procedure provided above. A Wald type confidence interval for $\theta$ may then be constructed as $\hat\theta_1\pm n^{-1/2}z_{1-\alpha/2}\hat\zeta$, where $1-\alpha$ is the confidence level, and $z_{1-\alpha/2}$ is the $(1-\alpha/2)$-quantile of the standard normality. The proof of these follows directly from standard theory for path-wise differentiable parameters. Interested readers are encouraged to consult the review by \cite{kennedy2022semiparametric}.

\section{A simulation study}
We conduct a simulation study to evaluate the finite-sample performance of the proposed estimation approach. The following data generating mechanism
is employed to simulate the data:

\begin{table}[H]
\setlength{\tabcolsep}{1pt}
\centering
\footnotesize
\label{tab:dgm}
\begin{tabular}{rl}
$X = (X_1, X_2, X_3)^\top$ & $\sim Be(2, 3)$ \\ 
$A | X $ & $ \sim Bern (\operatorname{logit}^{-1}(0.5 X_1 + 0.5 X_2 - 1))$ \\
$Z | A, X$ & $ \sim Bin (3, \operatorname{logit}^{-1}(-1.7 + 1.5 A + 0.5 X_3^2)$ \\
$M | Z, A, X$ & $ \sim Bin (3, \operatorname{logit}^{-1}(-1.5 + \lambda_1 Z + \lambda_2 A + 0.4 X_2 + 0.2 X_3))$ \\
$Y | M, Z, A, X$ & $\sim Bern(\operatorname{logit}^{-1}(0.4 M + \gamma_1 Z + \gamma_2 A - 0.5 \cos (X_1) - 1.5))$,
\end{tabular}
\label{tab:sim}
\end{table}
\noindent where $(\lambda_1, \lambda_2, \gamma_1, \gamma_2)^\top = (1.2 ,1.5, 1.2, 1.2)^\top$ are pre-specified parameters
; $Be(\alpha, \beta)$ denotes a Beta distribution with shape parameters $\alpha$ and $\beta$; $Bern(\pi)$ denotes a Bernoulli distribution with probability $\pi$; $Bin(n, \pi)$ denotes a binomial distribution with $n$ the number of independent trials and $\pi$ the success probability. 

Four different settings are considered. In the first two settings, we set $\lambda_1=0$ (setting 1) and $\lambda_2 = 0$ (setting 2) to assume away the direct effect of $Z$ on $M$ and of $A$ on $M$, so that no effect is mediated through path $P_3$ and $P_4$, respectively. In the two other settings, we set $\gamma_1=0$ (setting 3) and $\gamma_2=0$ (setting 4) to assume away the direct effect of $Z$ on $Y$ and of $A$ on $Y$, so that no effect is mediated through path $P_2$ and $P_1$, respectively. Across all settings, two sets of observed baseline covariates are considered, namely $X$ and $W = f(X)$, where $W = (W_1, W_2, W_3)^\top$ and
\[
W_1=  \exp (X_1 - 1), W_2 = (X_1 + X_2^2) / 4, W_3 = \sin (X_3),
\]
with the goal of assessing the performance of the estimators in settings where the nuisance models are misspecified.

In each setting, we calculate the true values of the path-specific effects $\psi_{P_1}$, $\psi_{P_2}$, $\psi_{P_3}$, $\psi_{P_4}$, and $\psi_{P_2 \vee P_3}$ based on equation (\ref{eq:iden}), then estimate these effects by the bias-corrected plug-in approach proposed in Section \ref{sec:estima}. To estimate the nuisance parameters, we used the Super Learner algorithm \citep{vanderLaanPolleyHubbard07}, which entails building an ensemble of regression algorightms that minimize the cross-validated risk. We used a sample mean, a generalized linear model, multivariate adaptive regression splines, and extreme gradient boosting in the ensemble \citep{friedman1991multivariate, chen2015xgboost}. The bias, $\sqrt{n}$-bias, standard deviation (SD) and coverage of the 95\% confidence interval of the proposed estimators are then reported. The simulation study is conducted 500 times with sample size $n \in \{500, 1000, 2000, 5000\}$. The R code for implementation can be found on \href{https://github.com/CI-NYC/recantingtwins_simulation}{https://github.com/CI-NYC/recantingtwins\_simulation}.

\begin{figure}
    \centering
    \includegraphics[scale = 0.45]{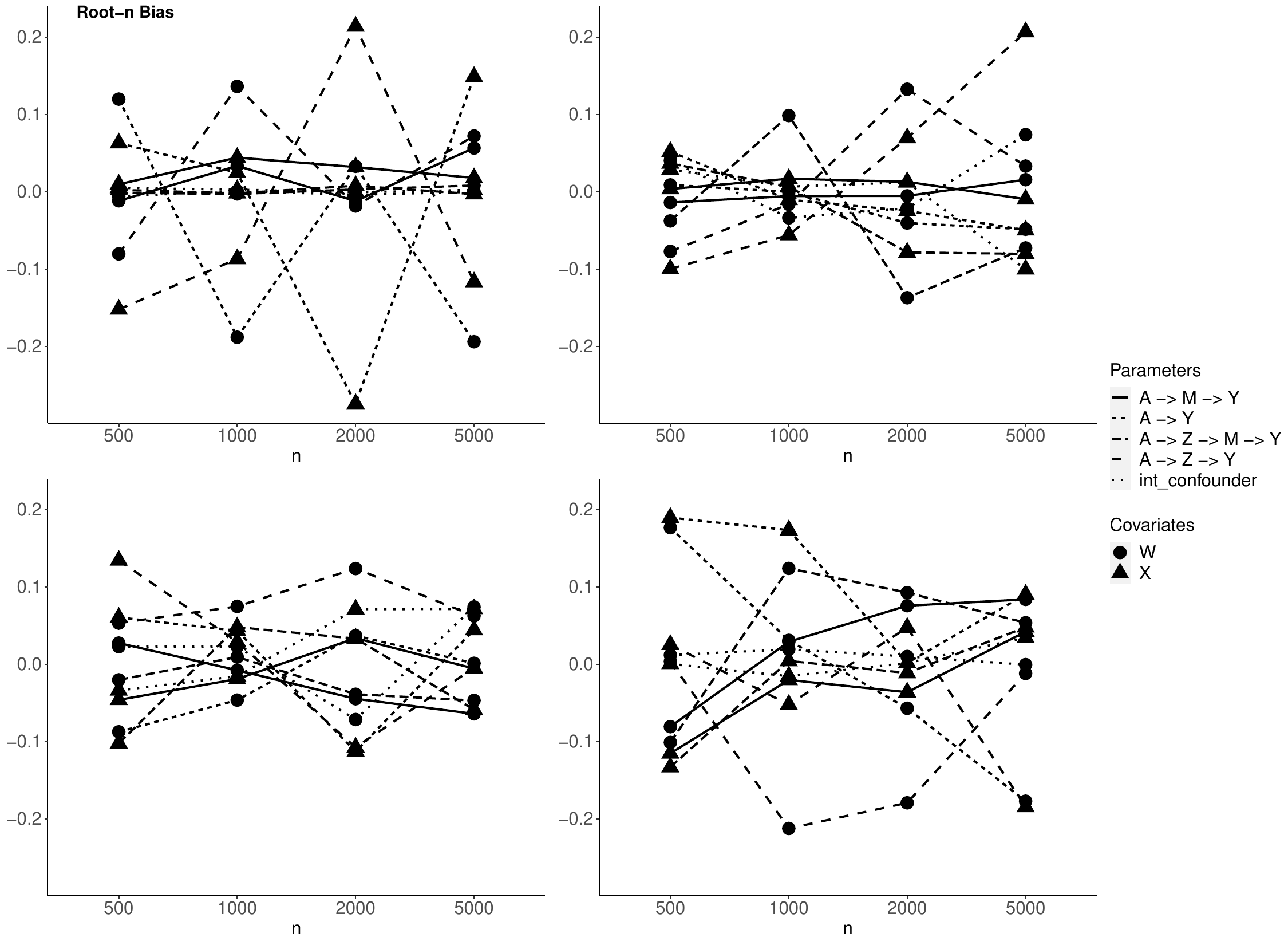}
    \caption{$\sqrt{n}$-bias for each setting. The top left figure corresponds to the case when $\lambda_1 = 0$, the top right figure corresponds to the case when $\lambda_2 = 0$, the bottom left figure corresponds to the case when $\gamma_1 = 0$, and the bottom right figure corresponds to the case when $\gamma_2 = 0$. $n$ means the sample size, each dot shape corresponds to one set of baseline covariates, each line type corresponds to the recanting-twins effects on a given path.}
    \label{fig:root_n_bias}
\end{figure}

\begin{figure}
    \centering
    \includegraphics[scale = 0.45]{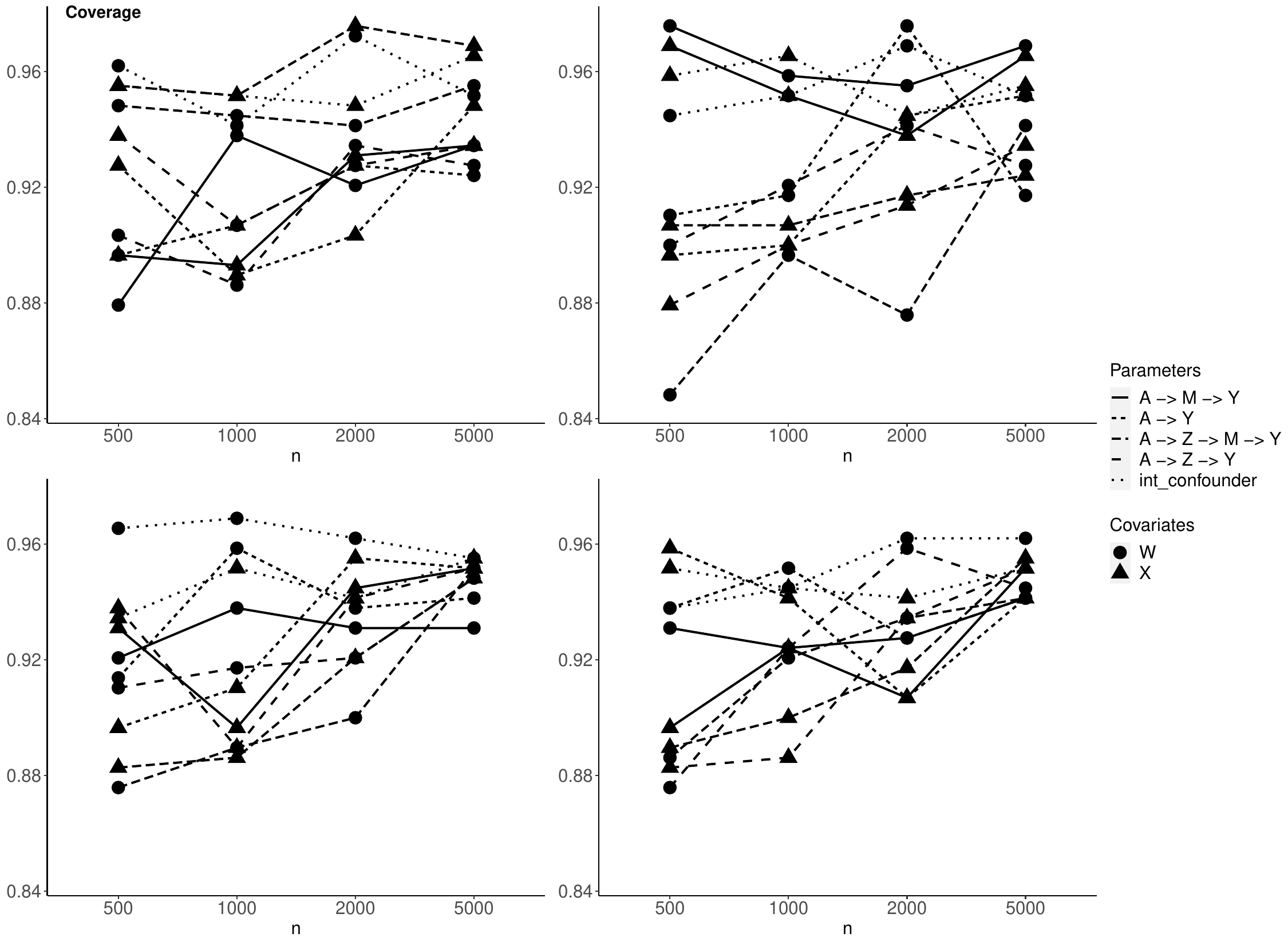}
    \caption{Coverage for each setting. The top left figure corresponds to the case when $\lambda_1 = 0$, the top right figure corresponds to the case when $\lambda_2 = 0$, the bottom left figure corresponds to the case when $\gamma_1 = 0$, and the bottom right figure corresponds to the case when $\gamma_2 = 0$. $n$ means the sample size, each dot shape corresponds to one set of baseline covariates, each line type corresponds to the recanting-twins effects on a given path.}
    \label{fig:coverage}
\end{figure}

Figure \ref{fig:root_n_bias} and \ref{fig:coverage} show the $\sqrt{n}$-bias and coverage of the 95\% confidence interval. Results for the bias and standard deviation can be found in the supplementary materials. Across all settings, the $\sqrt{n}$-bias of the estimators lies in a narrow range around zero, which confirms the properties of the estimator according to Theorem ~\ref{theo1}. The coverage of 95\% confidence intervals under small sample sizes is lower than nominal, but generally converges to the nominal coverage of $95\%$ under larger $n$, regardless of whether we use $X$ or $W$. 

\section{Illustrative application}
Over 20\% of US adults have a chronic pain condition \citep{rikard2023chronic}, which places them at increased risk of using and misusing opioids, and the attendant health consequences from such misuse \citep{mikosz2020indication,savych2019opioids,peters2018opioid, ozturk2021prescription, edlund_role_2013, glanz2019association,peirce_doctor_2012,rose2018potentially,cho2020risk,volkow2016opioid}. Previous work found that having a chronic pain condition increases risk of developing opioid use disorder (OUD) more than six-fold \citep{hoffman2023independent}. 
 
Much of this increased risk is suspected to be due to risky prescribing practices. 
For example, gabapentin and pregabalin, which are anti-epileptic drugs, are frequently prescribed off-label for non-shingles-related pain \citep{kuehn2022growing}. This off-label use has received recent attention as possibly problematic when co-prescribed with opioids \citep{kuehn2022growing,gomes2017gabapentin}. However, examining risky prescribing practices as an explanatory mechanism is complex, as there are many such prescribing practices and they may interact with one another within and across timepoints. In this illustrative example, we seek to limit this complexity by considering just one portion of the possible mechanism through gabapentin/pregabalin and opioid co-prescribing. 

We applied our proposed method to Medicaid claims data to understand: among those who had a chronic pain condition 
and an opioid prescription at the time of enrollment, to what extent did having an overlapping gabapentin/pregabalin prescription affect risk of incident opioid use disorder diagnosis? And to what extent did this total effect operate through mediators of new-onset anxiety or depressive disorder, and post-exposure confounders of benzodiazepine co-prescriptions and number of opioid providers?
We used Medicaid claims for non-pregnant, non-dual eligible adults who were enrolled for a minimum of 12 months, aged 35-64 years in 2016-2019 in the 26 states that enacted the Affordable Care Act in or prior to 2014. We excluded Maryland due to issues with their eligibility criteria data, which meant we included 25 states. Other exclusion/inclusion criteria are described in \citet{hoffman2023independent}. Our final cohort size was $N=36,732$ Medicaid beneficiaries.

We considered the following observed data $O=(W, A, Z, M, C, CY)$, where 
$C$ denotes whether the beneficiary remained enrolled in Medicaid by the outcome timepoint (1/0). We evaluated the outcome at 
24 months. We added inverse probability weights to the efficient influence function to incorporate $C$.

The variables $W$ and $A$ were measured during the first 6 months of Medicaid enrollment. We included the following baseline covariates: age in years; sex; race/ethnicity; English as their primary language; marriage/partnership status; household size; veteran status; 
income likely $>$133\% of the Federal Poverty Level; any inpatient or outpatient 
diagnosis of bipolar disorder, any anxiety disorder, attention deficit hyperactivity disorder (ADHD), any depressive disorder, or other mental disorder (e.g., anorexia, personality disorders); maximum daily dose of prescribed opioids; whether or not there was an overlapping opioid prescription with a: i) stimulant prescription, ii) benzodiazepine prescription, iii) muscle relaxant prescription; and whether or not opioid tapering occurred. 
Our treatment of interest, $A$, whether or not the individual had a gabapentin/pregabalin prescription that overlapped with their opioid prescription, was treated as a binary variable.

The variables $Z$ and $M$ were measured during the next 6 months of Medicaid enrollment, months 7-12. We considered the following set of post-treatment confounders, $Z$: whether or not the beneficiary had an overlapping 
benzodiazepine and opioid co-prescription (binary, 1/0) and number of opioid prescribers. We considered onset of any anxiety disorder or any depressive disorder in this same period as mediators.

Finally, the variables $C$ and $Y$ were measured during the following 12 months of Medicaid enrollment, months 13-24. Our censoring indicator was defined as $C=1$ if the beneficiary was still enrolled through month 24 and $C=0$ if the beneficiary disenrolled during this time. $Y$ was observed among those who had not disenrolled ($C=1$) and was defined using ICD-10 codes indicating 
opioid abuse or dependence, as has been done previously \citep{samples2018risk, samples2022psychosocial, hoffman2023independent}. 

Results are given in Table \ref{recantingtwins_illustration} below. We see that having an overlapping gabapentin and opioid prescription during the first 6 months of Medicaid enrollment increases risk of developing incident OUD by 1.6 percentage points (95\% CI: 1.19, 2.01 percentage points) by 24 months post-enrollment relative to having an opioid prescription without gabapentin among beneficiaries with a chronic pain condition and opioid prescription. This ATE can be decomposed into the paths shown in Table \ref{recantingtwins_illustration}. The majority of the effect is due to the direct effect of $A$ on $Y$, with the second-most contributing path being through $Z$ alone---number of opioid prescribers and having a benzodiazepine and opioid co-prescription. We see that almost none of the total effect operates through the hypothesized mediators of anxiety and depressive disorders.

\begin{table}[!t]
\centering
\begin{tabular}{lll}
\toprule
  Study Period  &  Path  &  Effects (95\% CI)   \\
\midrule
  24 months &  ATE   &      $0.0160 ~ (0.0119, 0.0201)$      \\
              &    $A \to Y$   &   $0.0117 ~ (0.0079, 0.0155)$             \\
               & $A \to Z \to Y$   &  $0.0041 ~ (0.0034, 0.0047)$  \\
 &  $A \to Z \to M \to Y$    &   $1.8 \times 10^{-5} ~ (-1.6 \times 10^{-5}, 5.2 \times 10^{-5})$             \\
              &   $A \to M \to Y$       &   $0.0002 ~ (3.0 \times 10^{-5}, 0.0005)$  \\
               &  Int Confounder  &  $7.7 \times 10^{-6} ~ (-2.3 \times 10^{-5}, 3.8 \times 10^{-5}) $\\
\bottomrule
\end{tabular}
\caption{Results summary, ATE stands for average treatment effect.}
\label{recantingtwins_illustration}
\end{table}

\section{Discussion}
In this paper, we formally and rigorously develop the proposal of  \citet{diaz2022causal} to address intermediate confounding in mediation analysis for decomposition of the average treatment effect into path-specific effects. Path-specific effects defined by recanting twins are identifiable from the observed data under different structural causal models that assume different levels of separability of the exposure and of the intermediate confounders. In the absence of intermediate confounding, recanting twin effects are equivalent to natural effects. However, when important intermediate confounders are present, a fine-grained decomposition of the total causal effect into multiple path-specific effects can  be obtained by our proposed approach. Recanting twin effects satisfy path-specific sharp null criteria, and can inform one about the presence of intermediate confounding (as well as its strength).

Many potential directions are available for future research. For instance, it is important to develop numerical methods to implement the bias-corrected plug-in estimator when the mediator and/or intermediate confounder are of high dimension (as is often the case in practice). One alternative estimation strategy is to develop structural nested models for recanting twin effects, as is done for natural path-specific effects. Finally, extensions of recanting twins to more complex settings are also needed, such as when multiple mediators or a repeatedly measured mediator is of interest. In those settings, there might be feedback relationships between the mediators and the intermediate confounders over time, which challenges the measurement of different path-specific effects and of the intermediate confounding effect.
\section*{Acknowledgments}
Iv\'an D\'iaz and Kara Rudolph were supported through a
Patient-Centered Outcomes Research Institute (PCORI) Project Program
Award (ME-2021C2-23636-IC) and through the National Institute on Drug Abuse (R01DA053243).

\section*{Conflict of interests}
All authors declare that they have no conflicts of interest.
\bibliographystyle{plainnat}
\bibliography{refs}
\newpage
\appendix
\section{Proofs}
\subsection{Equivalence of natural effects and recanting-twin effects in the absence of intermediate confounding (Lemma 1)}\label{app:a1}
    First, note that when $U\in \mathcal{U}_1$, i.e. $A$ does not cause $Z$, the natural effects via path $P_2$ and $P_3$ are null. Indeed,
    \begin{align*}
        Y_{S_1} = Y (0, Z(1), M (1, Z(1))) &= Y (0, Z(0), M (1, Z(1))) = Y_{S_2}\\
                &= Y (0, Z(0), M (1, Z(0))) = Y_{S_3}.
    \end{align*}
    So $Y_{S_1} - Y_{S_2} = Y_{S_2} - Y_{S_3}=0$. We also have:
    \begin{align*}
        Y'_{S_1} = Y (0, Z(1), M (1, T(1)))= Y (0, Z(0), M (1, T(1))) = Y'_{S_2}.
    \end{align*}
    So $Y'_{S_1}- Y'_{S_2}=0$. Besides, 
    \begin{align*}
        Y''_{S_2} = Y (0, T(0), M (1, Z(1)))
                = Y (0, T(0), M (1, Z(0))) = Y''_{S_3}
    \end{align*}
    So $Y''_{S_2}- Y''_{S_3}=0$. The recanting-twin path-specific effects via path $P_2$ and $P_3$ are thus also null when $U \in \mathcal{U}_1$.

    Next, when $U \in \mathcal{U}_2$, we have $M(z_1) = M(z_2)~\forall z_1, z_2$ almost surely. This implies that $Z$ does not cause $M$ for these individuals. As a result, $M(a,Z(a^*)) = M(a)$ for $a,a^*=0,1$. Hence,
    \begin{align*}
        Y_{S_1} - Y_{S_2} &= Y (0, Z(1), M (1)) - Y (0, Z(0), M (1))\\
        Y_{S_2} - Y_{S_3} &= Y (0, Z(0), M (1)) - Y (0, Z(0), M (1)) = 0.
    \end{align*}
    Besides, we also have $M(a,T(a^*)) = M(a)$ for $a,a^*=0,1$. Therefore,
    \begin{align*}
        Y'_{S_1} - Y'_{S_2} &= Y (0, Z(1), M (1)) - Y (0, Z(0), M (1))\\
        Y_{S_2} - Y_{S_3} &= Y (0, T(0), M (1)) - Y (0, T(0), M (1)) = 0.
    \end{align*}
    So the recanting-twin effect and the natural effect via path $P_2$ are non-null and equal, while the recanting-twin effect and the natural effect via path $P_3$ are both null.

    Finally, when $U \in \mathcal{U}_3$, we have $Y(m,z_1) = Y(m,z_0) ~\forall m,z_1, z_0$ almost surely. This implies that $Y(a,z_1,m) = Y(a,z_0,m)~\forall a,z_1, z_0,m$. Hence,
    \begin{align*}
        Y_{S_1} - Y_{S_2} &= Y (0, Z(1), M (1,Z(1)) - Y (0, Z(0), M (1,Z(1)) = 0\\
        Y_{S_2} - Y_{S_3} &= Y (0, M (1,Z(1))) - Y (0, M (1,Z(0))).
    \end{align*}
    It can also be shown that:
    \begin{align*}
        Y'_{S_1} - Y'_{S_2} &= Y (0, Z(1), M (1,T(1))) - Y (0, Z(0), M (1,T(1)))=0\\
        Y''_{S_2} - Y''_{S_3} &= Y (0, M (1,Z(1))) - Y (0, M (1,Z(0))).
    \end{align*}
     So the recanting-twin effect and the natural effect via path $P_2$ are both null, while the recanting-twin effect and the natural effect via path $P_3$ are non-null and equal. This finishes the proof.

\subsection{Proof of identification results}
\subsubsection{Identification under model $\mathcal{M}_1$}
\begin{align*}
    E\{Y_{S_j}'\mid W\} 
    &= \E\{Y(0,Z(a'),M(1,T(1)))\mid W \}\\
    &= \sum_{z,m,z'} \E\{Y(0,z,m)\mid Z(a')=z, M(1, z')=m, W\}\\ &~\P\{Z(a')=z\mid W\} ~\P\{M(1,z')=m\mid Z(a')=z, W\} ~\P\{Z(1)=z'\mid W \}\\
    &= \sum_{z,m,z'} \E\{Y(0,z,m)\mid A=0,z,m, W\} ~\P\{Z(a')=z\mid A=a', W\} ~\P\{M(1,z')\mid A=1,z',W\} \\
    &\qquad\qquad \P\{Z(1)=z'\mid A=1, W \}\\
    &=\sum_{z,m,z'} \E\{Y\mid A=0,z,m, W\} ~\P\{Z=z\mid A=a', W\} ~\P\{M=m\mid A=1,z',W\} \\
    &\qquad\qquad \P\{Z=z'\mid A=1, W \}\\
    &=\sum_{z,m} \E\{Y\mid A=0,z,m, W\} ~\P\{Z=z\mid A=a', W\} ~\P\{M=m\mid A=1,W\} 
\end{align*}
where $j=1,2$ and $a'=1$ when $j=1$ and $a'=0$ when $j=2$. The first equality results from assumption C2(i), the second one from assumption C1, and the third one from consistency.
\begin{align*}
    E\{Y_{S_j}''\mid W\} 
    &= \E\{Y(0,T(0),M(1,Z(a')))\mid W \}\\
    &= \sum_{z,m,z'} \E\{Y(0,z,m)\mid W\} ~\P\{Z(0)=z\mid W\} ~\P\{M(1,z')=m\mid W\} ~\P\{Z(a')=z'\mid W \}\\
    &= \sum_{z,m,z'} \E\{Y(0,z,m)\mid A=1,z,m, W\} ~\P\{Z(0)=z\mid A=0, W\} ~\P\{M(1,z')\mid A=1,z',W\} \\
    &\qquad\qquad \P\{Z(a')=z'\mid A=a', W \}\\
    &= \sum_{z,m,z'} \E\{Y\mid A=0,z,m, W\} ~\P\{Z=z\mid A=0, W\} ~\P\{M\mid A=1,z',W\} \\
    &\qquad\qquad \P\{Z=z'\mid A=a', W \}
\end{align*}
where $j=2,3$ and $a'=1$ when $j=2$ and $a'=0$ when $j=3$. The first equality results from assumption C2(i), and the second one from assumption C1. Next, when $j=2$, we have:
\begin{align*}
    E\{Y_{S_2}''\mid W\} 
    &= \sum_{z,m,z'} \E\{Y\mid A=0,z,m, W\} ~\P\{Z=z\mid A=0, W\} ~\P\{M\mid A=1,z',W\} \\
    &\qquad\qquad \P\{Z=z'\mid A=1, W \}\\
    &= \sum_{z,m,z'} \E\{Y\mid A=0,z,m, W\} ~\P\{Z=z\mid A=0, W\} ~\P\{M\mid A=1,W\}
\end{align*}    
And when $j=3$:
\begin{align*}
    E\{Y_{S_2}''\mid W\} 
    &= \sum_{z,m,z'} \E\{Y\mid A=0,z,m, W\} ~\P\{Z=z\mid A=0, W\} ~\P\{M\mid A=1,z',W\} \\
    &\qquad\qquad \P\{Z=z'\mid A=0, W \}
\end{align*}    
\subsubsection{Identification under model $\mathcal{M}_2$}    
\begin{align*}
    E\{Y_{S_j}'\mid W\} 
    &= \E\{Y(0,Z(a'),M(1,T(1)))\mid W \}\\
    &= \sum_{z,m,z'} \E\{Y(0,z,m)\mid W\} ~\P\{Z(a')=z; M(1,z')=m\mid W\} ~\P\{Z(1)=z'\mid W \}\\
    &= \sum_{z,m,z'} \E\{Y\mid A=0,z,m, W\} ~\P\{Z(a')=z; M(1,z')=m\mid W\} \\
    &\qquad\qquad \P\{Z=z'\mid A=1,W \}\\
    &= \sum_{z,m,z'} \E\{Y\mid A=0,z,m, W\} ~\P\{M(1,z')=m\mid W\} \\
    &\qquad\qquad \P\{Z(a')=z\mid W \}~\P\{Z=z'\mid A=1,W \}\\
    &= \sum_{z,m,z'} \E\{Y\mid A=0,z,m, W\} ~\P\{M=m\mid z',A=1,W\} \\
    &\qquad\qquad \P\{Z=z\mid A=a',W \}~\P\{Z=z'\mid A=1,W \}
\end{align*}
where $j=1,2$ and $a'=1$ when $j=1$ and $a'=0$ when $j=2$. The second equality results from assumption C6. The third equality results from assumption C4, and the fourth results from assumption C5.
\begin{align*}
    E\{Y_{S_j}''\mid W\} 
    &= \E\{Y(0,T(0),M(1,Z(a')))\mid W \}\\
    &= \sum_{z,m,z'} \E\{Y(0,z,m)\mid W\} ~\P\{Z(0)=z\mid W\} ~\P\{M(1,z')=m; Z(a')=z'\mid W \}\\
    &= \sum_{z,m,z'} \E\{Y(0,z,m)\mid A=1,z,m, W\} ~\P\{Z=z\mid A=0, W\} \\
    &\qquad\qquad \P\{M(1,z')=m\mid W\} ~\P\{Z(a')=z'\mid W \}\\
    &= \sum_{z,m,z'} \E\{Y\mid A=0,z,m, W\} ~\P\{Z=z\mid A=0, W\}  \\
    &\qquad\qquad \P\{M=m\mid A=1,z',W\}~ \P\{Z=z'\mid A=a', W \}\\
\end{align*}
where $j=2,3$ and $a'=1$ when $j=2$ and $a'=0$ when $j=3$. The second equality results from assumption C6. The third equality results from assumption C4 and C5, and the fourth results from consistency.
\subsubsection{Identification under model $\mathcal{M}_3$}
\begin{align*}
    E\{Y_{S_j}'\mid W\} 
    &= \E\{Y(1,Z_Y(a'),M(0,T_M(1)))\mid W \}\\
    &= \sum_{z_m} \E\{Y(0,Z_Y(a'),M(0,z_M))\mid W\}  ~\P\{Z_M(1)=z_M\mid W \}\\
    &= \sum_{z_m,z_y,m} \E\{Y(0,z_y,m)\mid W\} ~\P\{Z_Y(a')=z_y,M(0,z_M)=m \mid W\}  \\
    &\qquad\qquad\P\{Z_M(1)=z_M\mid W \}\\
    &= \sum_{z_m,z_y,m} \E\{Y(0,z_y,m)\mid W\} ~\P\{Z_Y(a')=z_y\mid W\} ~\P\{M(0,z_M)=m \mid W\}  \\
    &\qquad\qquad\P\{Z_M(1)=z_M\mid W \}\\
    &= \sum_{z_m,z_y,m} \E\{Y\mid A = 0, Z = z_Y, m, W\} ~\P\{Z=z_y\mid A = a', W\} \\
    &\qquad\qquad\P\{M=m \mid A = 0, Z = z_M, W\} ~\P\{Z=z_M\mid A = 1, W \}
\end{align*}
where $j=1,2$ and $a'=1$ when $j=1$ and $a'=0$ when $j=2$. The third equality results from assumption C8(ii). The forth equality results from assumption C9(ii), and the fifth results from assumption C7.
\begin{align*}
    E\{Y_{S_j}''\mid W\} 
    &= \E\{Y(1,T_Y(0),M(0,Z_M(a')))\mid W \}\\
    &= \sum_{z_y}\E\{Y(1,z_y,M(0,Z_M(a')))\mid W \}~\P\{Z_Y(0)=z_y\mid W\}\\
    &= \sum_{z_m,z_y,m}\E\{Y(1,z_y,m)\mid W \}~ \P\{ M(0,z_m) = m, Z_M(a') = z_m \mid W\}\\
    &\qquad\qquad \P\{Z_Y(0)=z_y\mid W\}\\
    &= \sum_{z_m,z_y,m}\E\{Y(1,z_y,m) \mid W \}~ \P\{ M(0,z_m) = m\mid W\} ~\\
    &\qquad\qquad \P\{ Z_M(a') = z_m \mid W\}~\P\{Z_Y(0)=z_y\mid W\}\\   
    &=\sum_{z_m,z_y,m}\E\{Y\mid A = 1, Z = z_y, m, W \}~ \P\{ M = m\mid A = 0, Z = z_m, W\} ~\\
    &\qquad\qquad \P\{ Z = z_m \mid A = a', W\}~\P\{Z=z_y\mid A = 0, W\}  
\end{align*}
where $j=2,3$ and $a'=1$ when $j=2$ and $a'=0$ when $j=3$. The third and forth equalities result from assumption C8i, and the fifth results from assumption C7.
\subsection{Efficient influence functions and second-order remainder terms}
\subsubsection{The proof for $\theta_1 = E\{Y_{S_1}\}$}
\begin{align} \theta_1 := E\{Y_{S_1}\}= \int E(Y\mid a^*,z,m,w)\cdot f_Z(z\mid m,a',w)\cdot f_M(m\mid a',w)\cdot f_W(w)~ dzdmdw
\end{align}
Assuming discrete $(Z,M,W)$ data, we have $\theta_1 = \sum_{z,m,w} \theta_{zmw}$, where
\begin{align*} \theta_{zmw} =\underbrace{E(Y\mid a^*,z,m,w)}_\text{$\theta_{zmw1}$} \cdot \underbrace{P(z\mid m,a',w)}_\text{$\theta_{zmw2}$}\cdot \underbrace{P(m\mid a',w)}_\text{$\theta_{zmw3}$} \cdot \underbrace{P(w)}_\text{$\theta_{zmw4}$}  \end{align*}
Denote $\Theta_{zmw} = \begin{pmatrix} \theta_{zmw1} & \theta_{zmw2} & \theta_{zmw3} & \theta_{zmw4}\end{pmatrix}^T$. By using the Delta method for functionals, one then has:
\[\If(\theta_1) = \sum_{zmw}\bigg[\frac{\partial\theta_{zmw}}{\partial\Theta_{lm}}\bigg]^T\If(\Theta_{zmw})\]
where IF$(\cdot)$ denotes the influence function. Besides,
\begin{align*}
  \bigg[\frac{\partial\theta_{zmw}}{\partial\Theta_{zmw}}\bigg]^T &= \begin{pmatrix} \theta_{zmw2}\theta_{zmw3}\theta_{zmw4} & \theta_{zmw1}\theta_{zmw3}\theta_{zmw4} & \theta_{zmw1}\theta_{zmw2}\theta_{zmw4} & \theta_{zmw1}\theta_{zmw2}\theta_{zmw3}\end{pmatrix}\\
  \If (\theta_{zmw1}) &= \frac{I(a^*,z,m,w)}{P(a^*,z,m,w)}\{Y-E(Y\mid a^*,z,m,w)\}\\
  \If (\theta_{zmw2}) &= \frac{I(m,a',w)}{P(m,a',w)}\{I(Z=z)-P(z\mid m,a',w)\}\\
  \If (\theta_{zmw3}) &= \frac{I(a',w)}{P(a',w)}\{I(M=m)-P(m\mid a',w)\}\\
  \If (\theta_{zmw4}) &= I(w)-P(w) 
\end{align*}
where $I(a^*,z,m,w)=I(A=a^*,Z=z,M=m,W=w)$ and so forth. As a result,
\begin{align*}
  \If(\theta_1) &= \sum_{z,m,w}~ I(a^*,z,m,w)\frac{P(z\mid m,a',w)P(m\mid a',w)P(w)}{P(a^*,z,m,w)}\{Y-E(Y\mid a^*,z,m,w)\}\\
                &\quad\quad+ I(m,a',w)\frac{E(Y\mid a^*,z,m,w)P(m\mid a',w)P(w)}{P(m,a',w)}\{I(Z=z)-P(z\mid m,a',w)\}\}\\
                &\quad\quad + I(a',w)\frac{E(Y\mid a^*,z,m,w)P(z\mid m,a',w)P(w)}{P(a',w)}\{I(M=m)-P(m\mid a',w)\}\\
                &\quad\quad + E(Y\mid a^*,z,m,w)P(z\mid m,a',w)P(m\mid a',w)\{I(w) - P(w)\}\vphantom{\frac{a}{b}}
\end{align*}
Some algebraic transformations then give:
\begin{align*}
  \If(\theta_1) &= \frac{I(a^*)}{P(a'\mid W)}~\frac{P(a'\mid Z,M,W)}{P(a^*\mid Z,M,W)}~\{Y-E(Y\mid a^*,Z,M,W)\} \\
                &+ \frac{I(a')}{P(a'\mid W)} \bigg[E(Y\mid a^*,Z,M,W) - E\big\{E(Y\mid a^*,Z,M,W) \mid  M,W,a'\big\}\bigg]\\
                &+ \frac{I(a')}{P(a'\mid W)}\bigg[E\big\{E(Y\mid a^*,Z,M,W) \mid  M,W,a'\big\} - E\{ E(Y\mid a^*,Z,M,W)\mid  a',W\}\bigg]\\
                &+E\{ E(Y\mid a^*,Z,M,W)\mid  a',W\} - \theta_1 \vphantom{\bigg[]}
\end{align*}
Removing the redundant terms, one then has:
\begin{align*}
  \If(\theta_1) &= \frac{I(a^*)}{P(a'\mid W)}~\frac{P(a'\mid Z,M,W)}{P(a^*\mid Z,M,W)}~\{Y-E(Y\mid a^*,Z,M,W)\} \\
                &+ \frac{I(a')}{P(a'\mid W)} E(Y\mid a^*,Z,M,W)
                  - \frac{I(a')}{P(a'\mid W)} E\{ E(Y\mid a^*,Z,M,W)\mid  a',W\}\\
                &+ E\{ E(Y\mid a^*,Z,M,W)\mid  a',W\} - \theta_1 \vphantom{\frac{a}{b}}
\end{align*}
We denote the following nuisance parameters:
\begin{align*}
  b(a,w) &= P(a\mid w)\\
  c(a,z,m,w) &= P(a\mid z,m,w)\\
  d(a,z,m,w) &= E(Y\mid a,z,m,w)\\
  e(a',w) &= E\big\{E(Y\mid a^*\mid Z,M,W)\mid a',w \big\}
\end{align*}
Note that $\theta_1(\eta) = E\{e(a',W)\}$. The remainder term can then be expressed as:
\begin{align*}
  R(\eta,\eta_1)&= E\bigg\{\frac{I(a^*)}{b_1(a',W)}~\frac{c_1(a',Z,M,W)}{c_1(a^*,Z,M,W)}~\{Y-d_1(a^*,Z,M,W)\}\bigg\} \\
                &+ E\bigg\{ \frac{I(a')}{b_1(a',W)} d_1(a^*,Z,M,W)
                  - \frac{I(a')}{b_1(a',W)} e_1(a',W)\bigg\}\\
                &+ \textcolor{red}{E\bigg\{e_1(a',W) - \theta_1(\eta_1) \vphantom{\frac{a}{b}}
                  \bigg\} + \theta_1(\eta_1) - \theta_1(\eta)}
\end{align*}
Consider the first component of $R(\eta,\eta_1)$:
\begin{align*}
  &E\bigg\{\frac{I(a^*)}{b_1(a',W)}~\frac{c_1(a',Z,M,W)}{c_1(a^*,Z,M,W)}~\big[Y-d_1(a^*,Z,M,W)\big]\bigg\} \\
  =&E\bigg\{\frac{c(a^*,Z,M,W)}{b_1(a',W)}~\frac{c_1(a',Z,M,W)}{c_1(a^*,Z,M,W)}~\big[d(a^*,Z,M,W)-d_1(a^*,Z,M,W)\big]\bigg\}\\
  =&\underbrace{E\bigg\{\frac{c_1(a',Z,M,W)}{b_1(a',W)}~\bigg[\frac{c(a^*,Z,M,W)}{c_1(a^*,Z,M,W)}-1\bigg]~\big[d(a^*,Z,M,W)-d_1(a^*,Z,M,W)\big]\bigg\}}_{R_1(\eta,\eta_1)} \\
  +&\textcolor{orange}{E\bigg\{\frac{c_1(a',Z,M,W)}{b_1(a',W)}~\big[d(a^*,Z,M,W)-d_1(a^*,Z,M,W)\big]
     \bigg\}}
\end{align*}
And the second component of $R(\eta,\eta_1)$:
\begin{align*}
  &E\bigg\{ \frac{I(a')}{b_1(a',W)} d_1(a^*,Z,M,W)
    - \frac{I(a')}{b_1(a',W)} e_1(a',W)\bigg\}\\
  =& E\bigg\{ \frac{I(a')}{b_1(a',W)} \big[d_1(a^*,Z,M,W)
     - d(a^*,Z,M,W)\big]\bigg\} + E\bigg\{ \frac{I(a')}{b_1(a',W)}\big[d(a^*,Z,M,W) - e_1(a',W)\big]\bigg\}\\
  =&\textcolor{blue}{E\bigg\{ \frac{c(a',Z,M,W)}{b_1(a',W)} \big[d_1(a^*,Z,M,W)
     - d(a^*,Z,M,W)\big]\bigg\}} \\
     &+ \textcolor{red}{E\bigg\{ \frac{I(a')}{b_1(a',W)}\big[E(d(a^*,Z,M,W)\mid a',W) - e_1(a',W)\big]\bigg\}}\\
\end{align*}
The sum of the two red terms equals:
\begin{align*}
  &E\{e_1(a',W) - e(a',W)\} +  E\bigg\{ \frac{b(a',W)}{b_1(a',W)}\big[e(a',W) - e_1(a',W)\big]\bigg\}\\
  = &\underbrace{E\bigg\{ \bigg[\frac{b(a',W)}{b_1(a',W)}-1\bigg]\big[e(a',W) - e_1(a',W)\big]\bigg\}}_{R_2(\eta,\eta_1)}
\end{align*}
The blue term can then be rewritten as:
\begin{align*}
  &\underbrace{E\bigg\{ \frac{c(a',Z,M,W)-c_1(a',Z,M,W)}{b_1(a',W)} \big[d_1(a^*,Z,M,W)
    - d(a^*,Z,M,W)\big]\bigg\}}_{R_3(\eta,\eta_1)} \\
  &+ \textcolor{orange}{E\bigg\{ \frac{c_1(a',Z,M,W)}{b_1(a',W)} \big[d_1(a^*,Z,M,W)
    - d(a^*,Z,M,W)\big]\bigg\}} 
\end{align*}
Notice that the sum of the two orange terms is zero. We thus have $R(\eta,\eta_1) = R_1(\eta,\eta_1) + R_2(\eta,\eta_1) + R_3(\eta,\eta_1)$. This finishes the proof for $\theta_1$.

\subsubsection{The proof for $\theta'_j = E\{Y_{S_j}'\}; j=1,2$}
Denote $\theta_{j}' = E\{Y_{S_j}'\}$ where $j=1,2$: 
\begin{align*}
  \theta_{j}' = E\bigg\{\int E(Y\mid a^*,z,m,W)~dP(z\mid a_j,W)~ dP(m\mid a',W) \bigg\}
\end{align*}
We have $a_j = a'$ when $j=1$; and $a_j = a^*$ when $j=2$. One then has:
\begin{align*}
  \If(\theta_{j}') &= \sum_{z,m,w}~ I(a^*,z,m,w)\frac{P(z\mid a_j,w)P(m\mid a',w)P(w)}{P(a^*,z,m,w)}\{Y-E(Y\mid a^*,z,m,w)\} \\
                   & \quad \quad + I(a_j,w)\frac{E(Y\mid a^*,z,m,w)P(m\mid a',w)P(w)}{P(a_j,w)}\{I(Z=z)-P(z\mid a_j,w)\}\\
                   & \quad \quad + I(a',w)\frac{E(Y\mid a^*,z,m,w)P(z\mid a_i,w)P(w)}{P(a',w)}\{I(M=m)-P(m\mid a',w)\}\\
                   &\quad\quad + E(Y\mid a^*,z,m,w)P(z\mid a_j,w)P(m\mid a',w)\{I(w) - P(w)\}\vphantom{\frac{a}{b}}
\end{align*}
We further denote:
\begin{align*}
  g(a) &= g(z,a,W) = P(z\mid a,W)\\
  h(a) &= h(m,a,W) = P(m\mid a,W)\\
  h^*(a) &= h^*(m,a,z,W) = P(m\mid a,z,W)\\
  b(a) &= b(a,W) = P(a\mid W)\\
  d(a) &= d(a,z,m,W) = E(Y\mid a,z,m,W)
\end{align*}
In other words, all nuisance parameters are functions of $a$ and of $(m,z,w)$, though we omit the latter to simplify the notations. One then has:
\begin{align*}
  \If(\theta_{j}') &= \sum_{z,m}~ I(a^*,z,m)\frac{g(a_j)~h(a')}{g(a^*)~h^*(a^*)~b(a^*)}[Y-d(a^*)] \\
                   & \quad \quad + I(a_j)\frac{d(a^*)~h(a')}{b(a_j)}\{I(Z=z)-g(a_j)\}\\
                   & \quad \quad + I(a')\frac{d(a^*)~g(a_j)}{b(a')}\{I(M=m)-h(a')\}\\
                   &\quad\quad + d(a^*)~g(a_j)~h(a') - \theta_j'\vphantom{\frac{a}{b}}
\end{align*}
Note that the sum over $z$ and $m$ turns into an integration in more general cases, when these variables are continuous. The remainder term can now be expressed as:
\begin{align*}
  R(\eta,\eta_1) &= E\bigg\{
                   \sum_{z,m}~ I(a^*,z,m)\frac{g_1(a_j)~h_1(a')}{g_1(a^*)~h_1^*(a^*)~b_1(a^*)}[Y-d_1(a^*)] \\
                 & \quad \quad + I(a_j)\frac{d_1(a^*)~h_1(a')}{b_1(a_j)}\{I(Z=z)-g_1(a_j)\}\\
                 & \quad \quad + I(a')\frac{d_1(a^*)~g_1(a_j)}{b_1(a')}\{I(M=m)-h_1(a')\}\\
                 &\quad\quad + \textcolor{red}{d_1(a^*)~g_1(a_j)~h_1(a') \vphantom{\frac{a}{b}}\bigg\} - \theta_j'(\eta)}
\end{align*}
where $\theta_j'(\eta) = E\big\{\sum_{z,m}d(a^*)~g(a_j)~h(a') \big\}$.
We analyze the first component of $R(\eta,\eta_1)$:
\begin{align*}
  &E\bigg\{
    \sum_{z,m}~ I(a^*,z,m)\frac{g_1(a_j)~h_1(a')}{g_1(a^*)~h_1^*(a^*)~b_1(a^*)}[Y-d_1(a^*)] \bigg\}\\
  =&E\bigg\{\sum_{z,m}~\frac{g_1(a_j)~h_1(a')}{g_1(a^*)~h_1^*(a^*)~b_1(a^*)}~[d(a^*)-d_1(a^*)]~b(a^*)~g(a^*)~h^*(a^*)\bigg\}
  \\
  =&\underbrace{E\bigg\{\sum_{z,m}~g_1(a_j)~h_1(a')~\bigg[\frac{g(a^*)~h^*(a^*)~b(a^*)}{g_1(a^*)~h_1^*(a^*)~b_1(a^*)}-1\bigg]~\big [d(a^*)-d_1(a^*)\big]~\bigg\}}_\text{$R_1(\eta,\eta_1)$}\\ &+ \textcolor{blue}{E\bigg\{\sum_{z,m}~g_1(a_j)~h_1(a') ~ [d(a^*)-d_1(a^*)]~\bigg\}}
\end{align*}
and the second component of $R(\eta,\eta_1)$:
\begin{align*}
  &E\bigg\{\sum_{z,m}~I(a_j)\frac{d_1(a^*)~h_1(a')}{b_1(a_j)}~[I(Z=z)-g_1(a_j)]\bigg\}\\
  =&E\bigg\{\sum_{z,m}~\frac{d_1(a^*)~h_1(a')}{b_1(a_j)}~[g(a_j)-g_1(a_j)]~b(a_j)\bigg\}\\
  =&\underbrace{E\bigg\{\sum_{z,m}~d_1(a^*)~h_1(a')~[g(a_j)-g_1(a_j)]~\bigg[\frac{b(a_j)}{b_1(a_j)}-1\bigg]\bigg\}}_\text{$R_2(\eta,\eta_1)$}\\
     &+ \textcolor{blue}{E\bigg\{\sum_{z,m}~d_1(a^*)~h_1(a')~\big[g(a_j)-g_1(a_j)\big]\bigg\}}
\end{align*}
and the third component of $R(\eta,\eta_1)$:
\begin{align*}
  &E\bigg\{\sum_{z,m}I(a')\frac{d_1(a^*)~g_1(a_j)}{b_1(a')}\{I(M=m)-h_1(a')\}\bigg\}\\
  =&E\bigg\{\sum_{z,m}b(a')~\frac{d_1(a^*)~g_1(a_j)}{b_1(a')}\{h(a')-h_1(a')\}\bigg\}\\
  =&\underbrace{E\bigg\{\sum_{z,m}d_1(a^*)~g_1(a_j)~\bigg[\frac{b(a')}{b_1(a')}-1\bigg]~[h(a')-h_1(a')]\bigg\}}_\text{$R_3(\eta,\eta_1)$} \\
     &+\textcolor{red}{E\bigg\{d_1(a^*)~g_1(a_j)~[h(a')-h_1(a')]\bigg\}\}}
\end{align*}
Note that the two red terms sum up to:
\begin{align*}
  &E\bigg\{\sum_{z,m} d_1(a^*)~g_1(a_j)~h(a') - d(a^*)~g(a_j)~h(a') \bigg\} \\
  =& E\bigg\{\sum_{z,m} \big[d_1(a^*)~g_1(a_j) - d(a^*)~g(a_j)\big]~h(a') \bigg\}\\
  = & \textcolor{blue}{E\bigg\{\sum_{z,m} \big[d_1(a^*) - d(a^*)\big]~g_1(a_j)~h(a')\bigg\} + E\bigg\{\sum_{z,m} h(a')~d(a^*)~\big[g_1(a_j)-g(a_j)\big]~ \bigg\}}
\end{align*}
Finally, notice that the sum of the blue terms, $R_4(\eta,\eta_1)$, is of second order. This finishes the proof for $\theta_j'$.

\subsubsection{The proof for $\theta_3''=E\{Y''_{S_3}\}$}
Denote $\theta_3''= E\bigg\{\int E(Y\mid a^*,z,m,W)~ dP(z\mid a^*,W)~ dP(m\mid a',z',W)~ dP(z'\mid a^*,W) \bigg\}
$, we have:
\begin{align*}
  \If(\theta_{3}'') &= \sum_{z,m,z',w} \frac{I(a^*,z,m,w)}{P(a^*,z,m,w)}[Y-E(Y\mid a^*,z,m,w)]P(z\mid a^*,W)~ P(m\mid a',z',W)~ P(z'\mid a^*,W)~P(w)\\
                    &\quad \quad\quad + \frac{I(a^*,w)}{P(a^*,w)}~[I(Z=z)-P(z\mid a^*,w)]~ E(Y\mid a^*,z,m,w)~P(m\mid a',z',w)~P(z'\mid a^*,w)~P(w)\\
                    &\quad \quad\quad + \frac{I(a',z',w)}{P(a',z',w)}[I(M=m)-P(m\mid a',z',w)]~ E(Y\mid a^*,z,m,w)~P(z\mid a^*,w)~P(z'\mid a^*,w)~P(w)\\
                    &\quad \quad\quad + \frac{I(a^*,w)}{P(a^*,w)}[I(Z=z') - P(z'\mid a^*,w)]~E(Y\mid a^*,z,m,w)~P(z\mid a^*,w)~P(m\mid a',z',w)P(w)\\
                    &\quad \quad\quad + [I(w)-P(w)]~E(Y\mid a^*,z,m,w)~P(z\mid a^*,w)~P(m\mid a',z',w)~P(z'\mid a^*,W) \vphantom{\frac{a}{b}}
\end{align*}
We further denote:
\begin{align*}
  g(a,z) &= g(z,a,W) = P(z\mid a,W)\\
  h^*(a,z) &= h^*(m,a,z,W) = P(m\mid a,z,W)\\
  b(a) &= b(a,W) = P(a\mid W)\\
  d(a,z) &= d(a,z,m,W) = E(Y\mid a,z,m,W)
\end{align*}
One then has:
\begin{align*}
  \If(\theta_{3}'') &= \sum_{z,m,z'} \frac{I(a^*,z,m)}{h^*(a^*,z)~b(a^*)}[Y-d(a^*,z)]~ h^*(a',z')~ g(a^*,z')\\
                    &\quad \quad\quad + \frac{I(a^*)}{b(a^*)}~[I(Z=z)-g(a^*,z)]~ d(a^*,z)~h^*(a',z')~g(a^*,z') \\
                    &\quad \quad\quad + \frac{I(a',z')}{g(a',z')~b(a')}[I(M=m)-h^*(a',z')]~ d(a^*,z)~g(a^*,z)~g(a^*,z')\\
                    &\quad \quad\quad + \frac{I(a^*)}{b(a^*)}[I(Z=z') - g(a^*,z')]~d(a^*,z)~g(a^*,z)~h^*(a',z')\\
                    &\quad \quad\quad + d(a^*,z)~g(a^*,z)~h^*(a',z')~g(a^*,z') - \theta_3'' \vphantom{\frac{a}{b}}
\end{align*}
Note that the sum over $z$ and $m$ turns into an integration in more general cases, when these variables are continuous. The remainder term can now be expressed as:
\begin{align*}
  R(\eta,\eta_1) &= E\bigg\{\sum_{z,m,z'} \frac{I(a^*,z,m)}{h^*_1(a^*,z)~b_1(a^*)}[Y-d_1(a^*,z)]~ h^*_1(a',z')~ g_1(a^*,z')\\
                 &\quad \quad\quad + \frac{I(a^*)}{b_1(a^*)}~[I(Z=z)-g_1(a^*,z)]~ d_1(a^*,z)~h^*_1(a',z')~g_1(a^*,z') \\
                 &\quad \quad\quad + \frac{I(a',z')}{g_1(a',z')~b_1(a')}[I(M=m)-h^*_1(a',z')]~ d_1(a^*,z)~g_1(a^*,z)~g_1(a^*,z')\\
                 &\quad \quad\quad + \frac{I(a^*)}{b_1(a^*)}[I(Z=z') - g_1(a^*,z')]~d_1(a^*,z)~g_1(a^*,z)~h^*_1(a',z')\\
                 &\quad \quad\quad + \textcolor{red}{d_1(a^*,z)~g_1(a^*,z)~h^*_1(a',z')~g_1(a^*,z')  \vphantom{\frac{a}{b}}\bigg\} - \theta_3''(\eta)}
\end{align*}
where $\theta_3''(\eta) = E\bigg\{\sum_{z,m,z'}d(a^*,z)~g(a^*,z)~h^*(a',z')~g(a^*,z') \bigg\}$. The first component of $R(\eta,\eta_1)$ equals to:
\begin{align*}
  &E\bigg\{\sum_{z,m,z'} \frac{I(a^*,z,m)}{h^*_1(a^*,z)~b_1(a^*)}[Y-d_1(a^*,z)]~ h^*_1(a',z')~ g_1(a^*,z')\bigg\}\\
  =&\underbrace{E\bigg\{\sum_{z,m,z'} \bigg[\frac{h^*(a^*,z)~b(a^*)}{h^*_1(a^*,z)~b_1(a^*)}-1][d(a^*,z)-d_1(a^*,z)]~g(a^*,z)~ h^*_1(a',z')~ g_1(a^*,z')\bigg\}}_{\text{$R_1(\eta,\eta_1)$}}\\
  + &\textcolor{blue}{E\bigg\{\sum_{z,m,z'} [d(a^*,z)-d_1(a^*,z)]~g(a^*,z)~ h^*_1(a',z')~ g_1(a^*,z')\bigg\}}
\end{align*}
The second component of $R(\eta,\eta_1)$ equals to:
\begin{align*}
  &E\bigg\{\sum_{z,m,z'}\frac{I(a^*)}{b_1(a^*)}~[I(Z=z)-g_1(a^*,z)]~ d_1(a^*,z)~h^*_1(a',z')~g_1(a^*,z')\bigg\}\\
  =&E\bigg\{\sum_{z,m,z'}\frac{b(a^*)}{b_1(a^*)}~[g(a^*,z)-g_1(a^*,z)]~ d_1(a^*,z)~h^*_1(a',z')~g_1(a^*,z')\bigg\}\\
  =&\underbrace{E\bigg\{\sum_{z,m,z'}\bigg[\frac{b(a^*)}{b_1(a^*)}-1\bigg]~[g(a^*,z)-g_1(a^*,z)]~ d_1(a^*,z)~h^*_1(a',z')~g_1(a^*,z')\bigg\}}_{\text{$R_2(\eta,\eta_1)$}}\\
  +& \textcolor{orange}{E\bigg\{\sum_{z,m,z'}[g(a^*,z)-g_1(a^*,z)]~ d_1(a^*,z)~h^*_1(a',z')~g_1(a^*,z')
     \bigg\}}
\end{align*}
The third component of $R(\eta,\eta_1)$ equals to:
\begin{align*}
  &E\bigg\{\sum_{z,m,z'}\frac{I(a',z')}{g_1(a',z')~b_1(a')}[I(M=m)-h^*_1(a',z')]~ d_1(a^*,z)~g_1(a^*,z)~g_1(a^*,z')\bigg\}\\
  =&E\bigg\{\sum_{z,m,z'}\frac{g(a',z')b(a')}{g_1(a',z')~b_1(a')}[h^*(a',z')-h^*_1(a',z')]~ d_1(a^*,z)~g_1(a^*,z)~g_1(a^*,z')\bigg\}\\
  =&\underbrace{E\bigg\{\sum_{z,m,z'}\bigg[\frac{g(a',z')b(a')}{g_1(a',z')~b_1(a')}-1\bigg][h^*(a',z')-h^*_1(a',z')]~ d_1(a^*,z)~g_1(a^*,z)~g_1(a^*,z')\bigg\}}_{\text{$R_3(\eta,\eta_1)$}}\\
  +&\textcolor{brown}{E\bigg\{\sum_{z,m,z'}[h^*(a',z')-h^*_1(a',z')]~ d_1(a^*,z)~g_1(a^*,z)~g_1(a^*,z')\bigg\}}
\end{align*}
The forth component of $R(\eta,\eta_1)$ equals to:
\begin{align*}
  &E\bigg\{\sum_{z,m,z'} 
    \frac{I(a^*)}{b_1(a^*)}[I(Z=z') - g_1(a^*,z')]~d_1(a^*,z)~g_1(a^*,z)~h^*_1(a',z')\bigg\}\\
  =&E\bigg\{\sum_{z,m,z'} 
     \frac{b(a^*)}{b_1(a^*)}[g(a^*,z') - g_1(a^*,z')]~d_1(a^*,z)~g_1(a^*,z)~h^*_1(a',z')\bigg\}\\
  =&\underbrace{E\bigg\{\sum_{z,m,z'} 
     \bigg[\frac{b(a^*)}{b_1(a^*)}-1\bigg][g(a^*,z') - g_1(a^*,z')]~d_1(a^*,z)~g_1(a^*,z)~h^*_1(a',z')\bigg\}}_{\text{$R_4(\eta,\eta_1)$}}\\
  +&\textcolor{red}{E\bigg\{\sum_{z,m,z'} [g(a^*,z') - g_1(a^*,z')]~d_1(a^*,z)~g_1(a^*,z)~h^*_1(a',z')\bigg\}}\\
\end{align*}
Notice that the sum of the two red terms equals to:
\begin{align*}
  &E\bigg\{\sum_{z,m,z'} g(a^*,z')~\bigg[d_1(a^*,z)~g_1(a^*,z)~h^*_1(a',z') - d(a^*,z)~g(a^*,z)~h^*(a',z')\bigg]\bigg\}\\
  =&\textcolor{blue}{E\bigg\{\sum_{z,m,z'} g(a^*,z')~\big[d_1(a^*,z)-d(a^*,z)\big]~g_1(a^*,z)~h^*_1(a',z')\bigg\}}\\
  +&\textcolor{orange}{E\bigg\{\sum_{z,m,z'} g(a^*,z')~d(a^*,z)~\big[g_1(a^*,z)-g(a^*,z)\big]~h^*_1(a',z')\bigg\}}\\
  +&\textcolor{brown}{E\bigg\{\sum_{z,m,z'} g(a^*,z')~d(a^*,z)~g(a^*,z)~\big[h^*_1(a',z') - h^*(a',z')\big]\bigg\}}
\end{align*}
Notice also that the sum $R_5(\eta,\eta_1)$ of the two blue terms, $R_6(\eta,\eta_1)$ of the two orange terms and $R_7(\eta,\eta_1)$ of the two brown terms are of second order. This finishes the proof for $\theta_3''$\\

\subsubsection{The proof for $\theta_3 = E\{Y_{S_3}\}$}
Denote $\theta_3 = E\{Y_{S_3}\} = E\bigg\{\int E(Y\mid a^*,z,m,w)~dP(z\mid a^*,W)~dP(m\mid a',z,W) \bigg\}$, we have:
\begin{align*}
  \If(\theta_3) &= \sum_{z,m,w} \frac{I(a^*,z,m,w)}{P(a^*,z,m,w)}~[Y - E(Y\mid a^*,z,m,w)]~P(z\mid a^*,w)~P(m\mid a',z,w)~P(w)\\
                &\quad\quad\quad + \frac{I(a^*,w)}{P(a^*,w)}~[I(Z=z)-P(z\mid a^*,w)]~E(Y\mid a^*,z,m,w)~P(m\mid a',z,w)~P(w)\\
                &\quad\quad\quad + \frac{I(a',z,w)}{P(a',z,w)}~[I(M=m)-P(m\mid a',z,w)]~E(Y\mid a^*,z,m,w)~P(z\mid a^*,w)~P(w)\\
                &\quad\quad\quad + [I(W=w)-P(w)]~E(Y\mid a^*,z,m,w)~P(z\mid a^*,w)~P(m\mid a',z,w)
\end{align*}
We further denote:
\begin{align*}
  g(a) &= g(z,a,W) = P(z\mid a,W)\\
  h^*(a) &= h^*(m,a,z,W) = P(m\mid a,z,W)\\
  b(a) &= b(a,W) = P(a\mid W)\\
  d(a) &= d(a,z,m,W) = E(Y\mid a,z,m,W)
\end{align*}
All nuisance parameters are thus functions of $a$ and of $(m,z,w)$, though we omit the latter to simplify the notations. One then has:
\begin{align*}
  \If(\theta_3) &= \sum_{z,m} \frac{I(a^*,z,m)}{h^*(a^*)~b(a^*)}~[Y - d(a^*)]~h^*(a')\\
                &\quad\quad\quad + \frac{I(a^*)}{b(a^*)}~[I(Z=z)-g(a^*)]~d(a^*)~h^*(a')\\
                &\quad\quad\quad + \frac{I(a',z)}{g(a')~b(a')}~[I(M=m)-h^*(a')]~d(a^*)~g(a^*)\\
                &\quad\quad\quad + d(a^*)~g(a^*)~h^*(a') - \theta_3 \vphantom{\frac{a}{b}}
\end{align*}
The remainder term can now be expressed as:
\begin{align*}
  R(\eta,\eta_1) &= E\bigg\{\sum_{z,m} \frac{I(a^*,z,m)}{h_1^*(a^*)~b_1(a^*)}~[Y - d_1(a^*)]~h_1^*(a')\\
                 &\quad\quad\quad + \frac{I(a^*)}{b_1(a^*)}~[I(Z=z)-g_1(a^*)]~d_1(a^*)~h_1^*(a')\\
                 &\quad\quad\quad + \frac{I(a',z)}{g_1(a')~b_1(a')}~[I(M=m)-h_1^*(a')]~d_1(a^*)~g_1(a^*)\\
                 &\quad\quad\quad + \textcolor{red}{d_1(a^*)~g_1(a^*)~h_1^*(a')\bigg\} - \theta_3(\eta)} \vphantom{\frac{a}{b}}
\end{align*}
Note that $\theta_3(\eta) = E\big\{\sum_{z,m}d_(a^*)~g(a^*)~h^*(a')\big\}$. The first component of $R(\eta,\eta_1)$ is equivalent to:
\begin{align*}
  &E\bigg\{\sum_{z,m} \frac{I(a^*,z,m)}{h_1^*(a^*)~b_1(a^*)}~[Y - d_1(a^*)]~h_1^*(a')\bigg\}\\
  =&E\bigg\{\sum_{z,m} \frac{h_1^*(a')}{h_1^*(a^*)~b_1(a^*)}~[d(a^*) - d_1(a^*)]~h^*(a^*)~g(a^*)~b(a^*)\bigg\}\\
  =&\underbrace{E\bigg\{\sum_{z,m} h_1^*(a')~g(a^*)\bigg[\frac{h^*(a^*)~b(a^*)}{h_1^*(a^*)~b_1(a^*)}-1\bigg]~[d(a^*) - d_1(a^*)]\bigg\}}_{\text{$R_1(\eta,\eta_1)$}} + \textcolor{blue}{E\bigg\{\sum_{z,m}h_1^*(a')~g(a^*)~[d(a^*) - d_1(a^*)]\bigg\}}
\end{align*}
The second component of $R(\eta,\eta_1)$ is equivalent to:
\begin{align*}
  &E\bigg\{\sum_{z,m} \frac{I(a^*)}{b_1(a^*)}~[I(Z=z)-g_1(a^*)]~d_1(a^*)~h_1^*(a') \bigg\}\\
  =&E\bigg\{\sum_{z,m} \frac{b(a^*)}{b_1(a^*)}~[g(a^*)-g_1(a^*)]~d_1(a^*)~h_1^*(a') \bigg\}\\
  =&\underbrace{E\bigg\{\sum_{z,m} \bigg[\frac{b(a^*)}{b_1(a^*)}-1\bigg]~[g(a^*)-g_1(a^*)]~d_1(a^*)~h_1^*(a') \bigg\}}_{\text{$R_2(\eta,\eta_1)$}}  + \textcolor{blue}{E\bigg\{\sum_{z,m} [g(a^*)-g_1(a^*)]~d_1(a^*)~h_1^*(a') \bigg\}}
\end{align*}
And the third component of $R(\eta,\eta_1)$ is equivalent to:
\begin{align*}
  &E\bigg\{\sum_{z,m} \frac{I(a',z)}{g_1(a')~b_1(a')}~[I(M=m)-h_1^*(a')]~d_1(a^*)~g_1(a^*) \bigg\}\\
  =&E\bigg\{\sum_{z,m} \frac{g(a')~b(a')}{g_1(a')~b_1(a')}~[h^*(a')-h_1^*(a')]~d_1(a^*)~g_1(a^*) \bigg\}\\
  =&\underbrace{E\bigg\{\sum_{z,m} \bigg[\frac{g(a')~b(a')}{g_1(a')~b_1(a')}-1\bigg]~[h^*(a')-h_1^*(a')]~d_1(a^*)~g_1(a^*) \bigg\}}_{\text{$R_3(\eta,\eta_1)$}}  + \textcolor{red}{ E\bigg\{\sum_{z,m} [h^*(a')-h_1^*(a')]~d_1(a^*)~g_1(a^*) \bigg\}}
\end{align*}
Note that the two red terms sum up to:
\begin{align*}
  &E\bigg\{\sum_{z,m}h^*(a')~d_1(a^*)~g_1(a^*) - d(a^*)~g(a^*)~h^*(a')\bigg\} \\
  =& \textcolor{blue}{E\bigg\{\sum_{z,m} h^*(a')~[d_1(a^*)-d(a^*)]~g_1(a^*)\bigg\} + E\bigg\{\sum_{z,m} d(a^*)~[g_1(a^*)-g(a^*)]~h^*(a')\bigg\}}  
\end{align*}
Notice that the sum of the blue terms, i.e. $R_4(\eta,\eta_1)$ is of second order. This finishes the proof for $\theta_3$.

\subsection{Additional Results for Numerical Experiments}
\begin{figure}[tbh!]
    \centering
    \includegraphics[scale = 0.45]{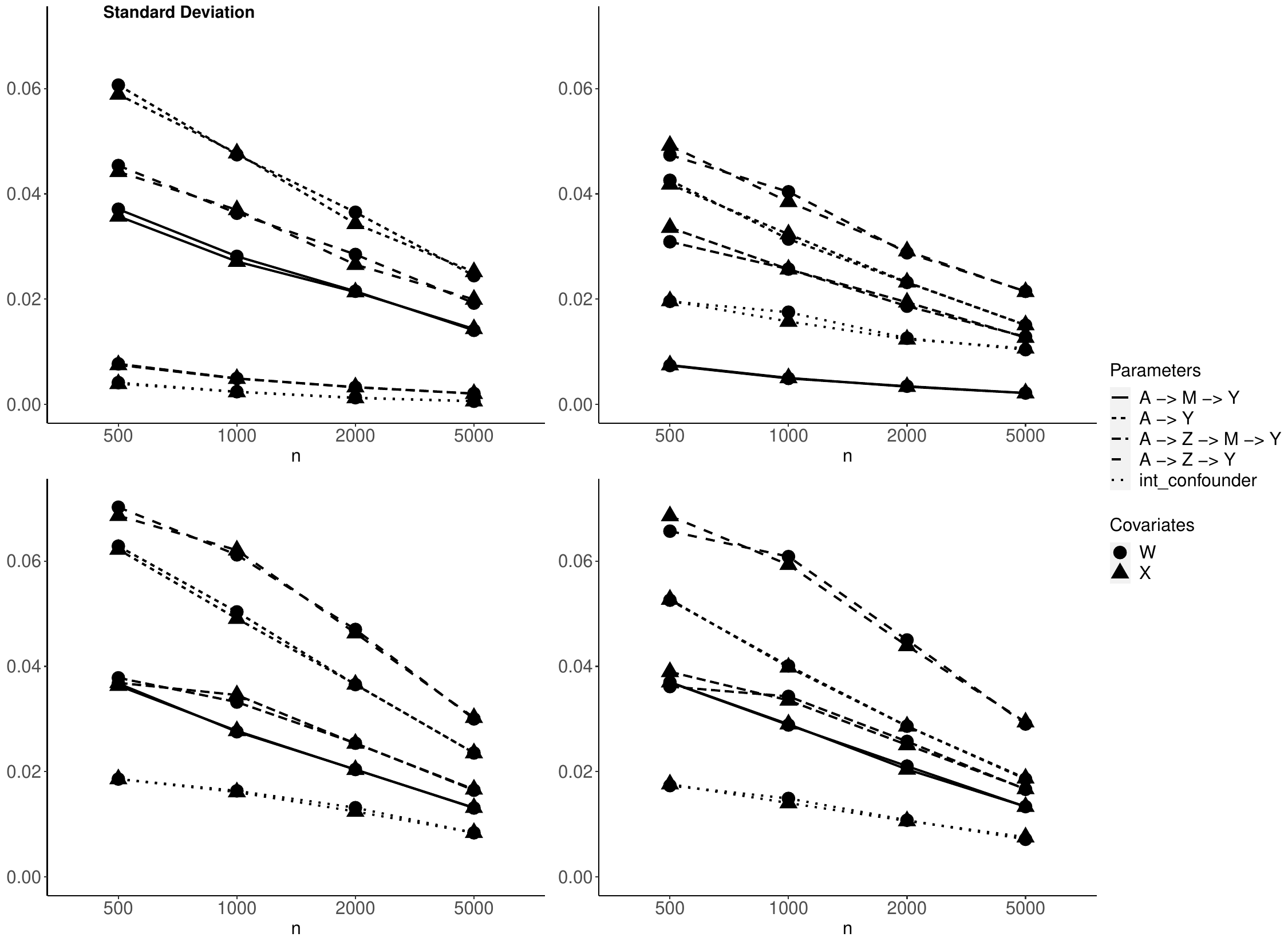}
    \caption{Standard deviation for each setting. The top left figure corresponds to the case when $\lambda_1 = 0$, the top right figure corresponds to the case when $\lambda_2 = 0$, the bottom left figure corresponds to the case when $\gamma_1 = 0$, and the bottom right figure corresponds to the case when $\gamma_2 = 0$. $n$ means the sample size, each dot shape corresponds to one set of baseline covariates, each line type corresponds to the recanting-twins effects on a given path.}
    \label{fig:standard_deviation}
\end{figure}

\begin{figure}
    \centering
    \includegraphics[scale = 0.45]{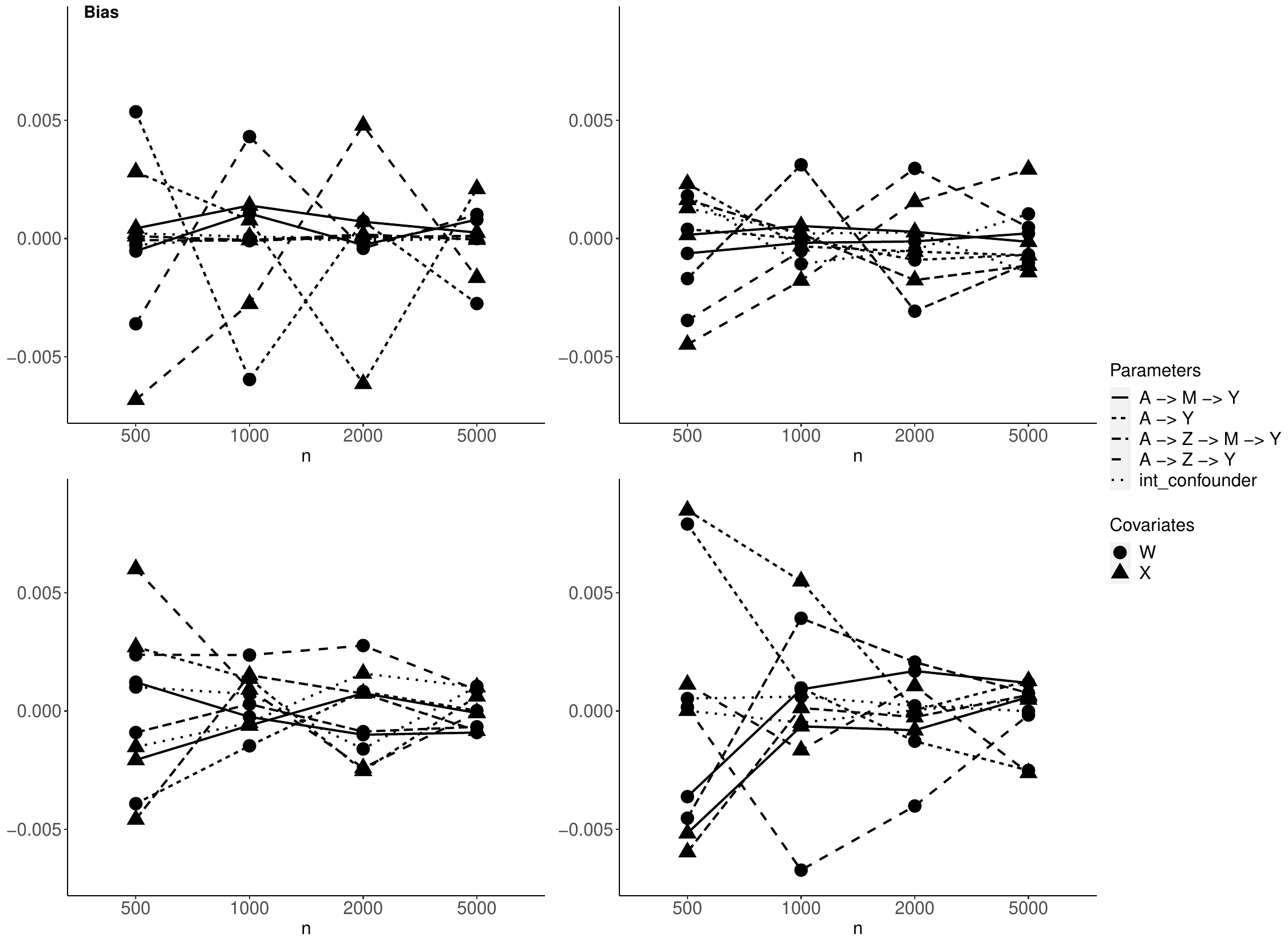}
    \caption{Bias for each setting. The top left figure corresponds to the case when $\lambda_1 = 0$, the top right figure corresponds to the case when $\lambda_2 = 0$, the bottom left figure corresponds to the case when $\gamma_1 = 0$, and the bottom right figure corresponds to the case when $\gamma_2 = 0$. $n$ means the sample size, each dot shape corresponds to one set of baseline covariates, each line type corresponds to the recanting-twins effects on a given path.}
    \label{fig:bias}
\end{figure}

\end{document}

%% file: _preamble.tex
\usepackage[inline,shortlabels]{enumitem}
\usepackage{float}
\usepackage{mathtools}
\usepackage{subdepth}
\usepackage{graphicx}
\usepackage[round]{natbib}
\usepackage[margin=1in]{geometry}
\usepackage{tikz}
\usetikzlibrary{arrows.meta}
\usetikzlibrary{decorations.markings}
\usepackage[linesnumbered]{algorithm2e}
\RestyleAlgo{ruled}

\tikzset{+ /.tip = {Bar[sep=-3pt 2,width=3pt 4]_[sep=0]}}
\usepackage[english]{babel}
\usepackage{longtable}
\usepackage{color}
\usepackage{mathtools}
\usepackage{amssymb,amsmath,amsthm}
\usepackage{multirow}
\usepackage[titletoc,title]{appendix}
\usepackage{authblk}
\usepackage{bbm}
\usepackage{setspace}
\usepackage{dsfont}
\usepackage[OT1]{fontenc}
\usepackage{subcaption}
\usepackage{refcount}
\usepackage{booktabs}
\graphicspath{ {../simulations/01_init_manuscript/graphs/} }

\usepackage{xr}
\usepackage[colorlinks,citecolor=blue,urlcolor=blue]{hyperref}

\usepackage[inline]{trackchanges}
\addeditor{Ivan}

\newtheorem{theorem}{Theorem}
\AtEndDocument{\refstepcounter{theorem}\label{finalthm}}
\AtEndDocument{\refstepcounter{equation}\label{finaleq}}

{
  \theoremstyle{definition}
  \newtheorem{assumption}{}
}
{
  \theoremstyle{definition}
  
}
{
  \theoremstyle{definition}
  
}

\renewcommand{\If}{\mbox{IF}}

\newtheorem{lemma}{Lemma}
\AtEndDocument{\refstepcounter{lemma}\label{finallemma}}

\newtheorem{definition}{Definition}

\renewcommand{\P}{\mathsf{P}}

\newcommand{\indep}{\mbox{$\perp\!\!\!\perp$}}

\newcommand{\E}{\mathsf{E}}
\renewcommand{\P}{\mathsf{P}}

\usepackage{accents}

\DeclarePairedDelimiterX{\norm}[1]{\lVert}{\rVert}{#1}

\pgfdeclarelayer{background}
\pgfsetlayers{background,main}
\usetikzlibrary{arrows,positioning}
\tikzset{
>=stealth',
punkt/.style={
rectangle,
rounded corners,
draw=black, very thick,
text width=6.5em,
minimum height=2em,
text centered},
pil/.style={
->,
thick,
shorten <=2pt,
shorten >=2pt,}
}
\newcommand{\Vertex}[3]
{\node[minimum width=0.6cm,inner sep=0.05cm] (#2) at (#1) {#3};
}
\newcommand{\Vertexr}[3]
{\node[rectangle, draw, minimum width=0.6cm,inner sep=0.05cm] (#2) at (#1) {#2};
}
\newcommand{\ArrowR}[3]%
{ \begin{pgfonlayer}{background}
\draw[->,#3] (#1) to[bend right=30] (#2);
\end{pgfonlayer}
}
\newcommand{\ArrowLW}[3]%
{ \begin{pgfonlayer}{background}
\draw[->,#3] (#1) to[bend left=30] (#2);
\end{pgfonlayer}
}

\newcommand{\ArrowL}[3]%
{ \begin{pgfonlayer}{background}
    \draw[->,#3] (#1) to[bend left=45] (#2);
  \end{pgfonlayer}
}

\newcommand{\EdgeL}[3]%
{ \begin{pgfonlayer}{background}
\draw[dashed,#3] (#1) to[bend right=-45] (#2);
\end{pgfonlayer}
}

\newcommand{\Arrow}[3]%
{ \begin{pgfonlayer}{background}
\draw[->,#3] (#1) -- +(#2);
\end{pgfonlayer}
}

\allowdisplaybreaks
\pgfarrowsdeclare{arcs}{arcs}{...}
{
  \pgfsetdash{}{0pt} 
  \pgfsetroundjoin   
  \pgfsetroundcap    
  \pgfpathmoveto{\pgfpoint{-5pt}{5pt}}
  \pgfpatharc{180}{270}{5pt}
  \pgfpatharc{90}{180}{5pt}
  \pgfusepathqstroke
}
\newcommand{\ArrowB}[3]%
{ \begin{pgfonlayer}{background}
    \draw[|-arcs,line width=0.4mm,shorten <= 0.3cm,shorten >= 0.3cm,#3] (#1) -- +(#2);
  \end{pgfonlayer}
}

\newcommand{\titlepaper}{Recanting twins: addressing intermediate
  confounding in mediation analysis}

\date{\today}

\author[1]{Tat-Thang Vo}
\author[2]{Nicholas Williams}
\author[3]{Richard Liu}
\author[2]{Kara E. Rudolph}
\author[3]{Iv\'an D\'iaz}
\affil[1]{\small Research Group EPIDERME, Faculty of Medicine, University Paris Est Creteil, France}
\affil[2]{\small Department of Epidemiology, Mailman School of Public Health, Columbia University, USA}
\affil[3]{\small Division of Biostatistics, Department of Population
  Health, New York University Grossman School of Medicine, USA}